\documentclass[preprint,preprintnumbers,amsmath,amssymb,nofootinbib]{revtex4}
\usepackage{here}
\usepackage{braket}
\usepackage{graphicx,color}

\newcommand\nn{\nonumber \\}

\newcommand\bx{{\bf x}}

\newcommand\br{{\bf r}}

\newcommand{\lk}{\left(}
\newcommand{\rk}{\right)}
\newcommand{\ltk}{\left\{}
\newcommand{\rtk}{\right\}}
\newcommand{\ldk}{\left[}

\newcommand{\rdk}{\right]}
\newcommand\beq{ \begin{eqnarray} }
\newcommand\eeq{ \end{eqnarray} }

\begin{document}

\title{Bose polaron in spherically symmetric trap potentials: \\
Ground states with zero and lower angular momenta}
\author{Kano Watanabe }
\email{b17m6c17m@kochi-u.ac.jp}
\affiliation{Department of Physics, Kochi University, 
Kochi 780-8520, Japan}
\author{Eiji Nakano}
\email{e.nakano@kochi-u.ac.jp}
\affiliation{Department of Physics, Kochi University, 
Kochi 780-8520, Japan}
\author{Hiroyuki Yabu} 
\email{yabu@se.ritsumei.ac.jp}
\affiliation{Department of Physics,
Ritsumeikan University, Kusatsu 525-8577, Siga, Japan}

\date{\today}

\begin{abstract}
Single-atomic impurities immersed in a dilute Bose gas 
in the spherically symmetric harmonic trap potentials are studied 
at zero temperature. 
In order to find the ground state of the polarons, 
we present a conditional variational method 
with fixed expectation values 
of the total angular momentum operators, 
$\hat{J}^2$ and $\hat{J}_z$, of the system, 
using a cranking gauge-transformation for bosons 
to move them in the frame co-rotating with the impurity.
In the formulation, 
the expectation value $\langle \hat{J^2}\rangle$ is shown to be shared 
in impurity and bosons, 
but the value $\langle \hat{J}_z\rangle$ is carried by the impurity due to the rotational symmetry. 
We also analyze the ground-state properties numerically obtained in this variational method 
for the system of the attractive impurity-boson interaction, 
and find that excited boson distributions around the impurity 
overlap largely with impurity's wave function
in their quantum-number spaces and also in the real space 
because of the attractive interaction employed. 
\end{abstract}


\maketitle


\section{Introduction}
Recently much attention has been devoted to atomic impurities 
embedded in ultra-cold atomic media 
because of experimental accessibility of such systems in controlled ways
and of observations of various kinds of quasi-particle properties of the impurities
: bosonic \cite{Catani1,Scelle1,Hohmann1,Compagno1,Jrgensen1,Hu1,Rentrop1} 
and fermionic \cite{Froehlich1,Scazza1,Cetina1} ones.
For instance, the coupling between impurities and medium  
can be tuned using the Fano-Feshbach resonances between atomic hyperfine states, 
and the spatial dimensionality or periodicity of the system can also be designed 
using the effects of external electromagnetic fields \cite{PethickSmith1,Pitaevskii1}. 
The quasi-particle energy, 
width and spectral weight of the impurity can be measured 
in radiofrequency spectroscopy \cite{Jrgensen1,Hu1,Froehlich1}, 
and the fine energy splitting of a trapped impurity be measured in the Ramsey spectroscopy 
with oscillating fields \cite{Rentrop1}. 
Also, the dynamical aspects of the polaron formation can be observed 
experimentally \cite{Cetina1}. 

Theoretical studies of such systems have been actively performed 
prior to the experiments and revealed 
that properties of impurity are diverse depending on 
the impurity-medium interaction and medium properties: 
When the medium is a Bose-Einstein condensate (BEC), 
the impurity interacting through the Bogoliubov phonon of the medium is called 
a Bose polaron \cite{Cucchietti1,Sacha1,Tempere1,Casteels1,Rath1,Shashi1,Li1,
Levinsen2,Dehkharghani5,Ardila2,Christensen1,Vlietinck1,
Grusdt1,Grusdt3,Shchadilova1,Shchadilova2,Grusdt6,Ashida1,Ardila6,Nielsen3,Mistakidis8,Mistakidis9} 
in analogy with that in electron-phonon systems 
\cite{polaronreview1,Landau1,FPZ1}, 
where the atomic impurity is a quasiparticle dressed with a virtual cloud of excited phonons. 
In the case of a degenerate Fermi-gas medium, 
the impurity is called Fermi polaron 
\cite{Chevy1,Massignan1,Schirotzek1,Ku1,Schmidt1,Kohstall1,Koschorreck1,Schmidt6, 
Vlietinck2,Trefzger1,Trefzger2,Massignan5,Lan1,Lan2,Nur1,Yi1,Parish5, 
Levinsen4,Kamikado1,Kain1,Schmidt5,Mistakidis10}, 
which is dressed with particle-hole excitations around the Fermi surface. 
In these studies 
the low energy $s$-wave contact interaction has been frequently used for the impurity-medium interaction.  
Other kinds of atomic polarons are also studied 
with unconventional impurity-medium interactions, e.g., 
$p$-wave interactions \cite{Levinsen5,Deng1}, 
dipolar-dipolar interactions \cite{Kain3,Ardila4}.
In the case of the impurity-medium coupling tuned around the unitarity limit, 
the impurity and medium atoms can form few-body bound states in the medium  \cite{Levinsen2,Shchadilova2,Levinsen4}, 
and consequently the quasi-particle residue almost vanishes. 
The above mentioned studies entirely assume zero temperature, 
but recently thermal evolutions of polarons have been investigated, 
where the medium temperature varies from cold degenerate to hot Boltzmann regimes 
for Fermi polarons \cite{Tajima1,Yan8,Tajima2}, 
and, for Bose polarons, the temperature varies over the BEC critical temperature \cite{Levinsen8,Guenther8}.

In many studies of the polaron that have been done so far,
the system is assumed to be spatially uniform, 
while the real experiments of the ultra-cold gas are usually done on the systems trapped 
in the harmonic potentials. 
In the present study, 
we consider a Bose polaron in spherically symmetric trap in three dimensions, 
where the angular momentum of the polaron gives the conserved quantum numbers 
instead of spacial momenta in the uniform system. 
In particular, 
we calculate the ground-state energy of a trapped Bose polaron 
of fixed total angular momentum, 
and make clear the distributions of the angular-momentum 
and other quantum numbers of the polaron between the impurity and excited bosons in medium. 
For this purpose, 
we develop a variational method 
with the fixed expectation value of the angular momentum operators. 

In Sec.~II,
we set up our system, and derive a Fr{\"o}hlich type effective Hamiltonian. 
In Sec.~III, 
we introduce a cranking gauge transformation, 
by which all bosons in medium are cranked to move in the co-rotating frame of impurity. 
In Sec.~IV, 
we develop a variational method to obtain the energy functional for the cranked Hamiltonian, 
and present variational solutions and distribution functions of the excited bosons. 
In Sec.~V, numerical results and discussion for them are shown. 
Sec.~VI is for the summary and outlook. 

\section{Fr{\"o}hlich type effective Hamiltonian}
We consider the system of a single atomic impurity interacting with a dilute Bose gas, 
where the impurity and the gas are trapped 
in the spherically-symmetric harmonic potentials with the same centers.
The impurity and bosons are all spinless, 
so that the total orbital angular momentum of the system is conserved.  
We also suppose that bosons are non-interacting, 
while the impurity-boson interaction is tuned finite by the Feshbach resonance method.  
Thus, 
when the interaction is turned off, 
all medium bosons occupy the lowest energy state of the trap potential to form a $T=0$ BEC. 
This system is described by the following effective Hamiltonian:
\beq
\mathcal{H}(\br)
&=& 
H_{ho}(\br)
+\int_{\br'} \, \phi^\dagger(\br')
\ldk 
-\frac{\hbar^2 {\nabla'}^2}{2 m_b} +\frac{m_b \omega_b^2}{2}{r'}^2 
\rdk 
\phi(\br')
\nn
&&+g \int_{\br'} \, \phi^\dagger(\br')
\delta^{(3)}(\br-\br') 
\phi(\br') 
\nn 
&=& 
H_{ho}(\br) 
+\sum_{s} E_{s}^{b}b_s^\dagger b_s 
+g \sum_{s,s'} \phi_s^{b*}(\br) \phi_{s'}^b(\br)
b_{s}^\dagger b_{s'} 
\label{FH1}
\eeq
where the freedoms of the impurity and medium boson are represented 
in the first and the second quantized form.
We have used the abbreviated notation for the spacial integral: 
$\int_{\br}\equiv \int {\rm d}\br^3$.  
The first term $H_{ho}(\br)$ is the Hamiltonian of the impurity 
trapped in the harmonic-oscillator potential: 
\beq
H_{ho}(\br) 
&=& 
-\frac{\hbar^2}{2m_I} \frac{1}{r^2} \partial_r\lk r^2 \partial_r \rk
+\frac{\hbar^2{\hat{L}}^2}{2m_I\, r^2}
+\frac{m_I \omega_I^2}{2}r^2 
\eeq
where $(r, \theta, \varphi)$ is the spherical coordinate of the impurity, 
and $m_I$ and $\omega_I$ is the impurity mass and the angular frequency of the trap. 
The squared orbital angular-momentum operator ${L}^2$ 
is represented by,
\begin{equation*}
{\hat{L}}^2=-\ldk \frac{1}{\sin\theta} \partial_\theta \lk \sin\theta \partial_\theta \rk
+\frac{1}{\sin^2\theta} \partial_\varphi^2 \rdk,
\end{equation*}
The second term in (\ref{FH1}) represents the Hamiltonian of the medium boson; 
the $m_b$ and $\omega_b$ are the mass and the trap angular frequency of the medium boson, 
and the coupling constant $g$ of the impurity-medium contact interaction is given 
by the s-wave scattering length $a$ as $g=2\pi a/m_r$ in low-temperature approximation.
The second line of (\ref{FH1}) is obtained 
with the substitution of the field operator expansion:
$\phi(\br')=\sum_s \phi_s^b(\br') b_s$ 
where the $\phi_s^b({\bf r})$ are the eigenfunctions 
of the harmonic oscillator potential for the eigenvalues $E^b_s$, 
and the $b_s$ and $b_s^\dagger$ are the corresponding bosonic annihilation and creation operators. 
The label $s$ representing the medium-boson states is 
the abbreviated notation for $s=( n, l, m )$: 
the principal, the azimuthal, and the magnetic quantum numbers, respectively. 

The explicit form of the Harmonic-oscillator eigenfunctions $\phi^\alpha_s({\bf r})$ 
for the medium boson ($\alpha=b$) and the impurity ($\alpha=I$) are denoted as,
\begin{equation}
     \phi_s^{\alpha}({\bf r}) =R^{\alpha}_{n l}(r) Y_{lm}(\theta, \varphi), 
\label{eq:koyukansu}
\end{equation}
where the angular part $Y_{lm}$ is the spherical harmonic $Y$-function,
and the radial part $R^{\alpha}_{nl}(r)$ are,
\begin{align}
R^{\alpha}_{nl}(r) &=
{\mathcal N}_{n,l}
\lk m_\alpha \omega_\alpha\rk^{\frac{3}{4}} 
\, \lk \sqrt{m_\alpha \omega_\alpha} r\rk^{l}\, 
e^{-\frac{m_{\alpha} \omega_{\alpha}}{2}r^2}
L^{\lk \frac{1+2l}{2}\rk}_n\lk m_\alpha \omega_\alpha r^2 \rk, 
\label{eq:koyukansu2yabu}\\
{\mathcal N}_{nl}&=\sqrt{\frac{2^{n+l+2} n!}{\sqrt{\pi} \lk 2n+2l+1 \rk !!} }.
\label{eq:normalization}
\end{align}
The Laguerre function $L_n^{(k)}(x)$ that we use in this paper is defined by
\begin{equation*}
     L_n^{(k)}(x) =\frac{e^x x^{-k}}{n!} \frac{{\rm d}^n}{{\rm d}x^n}\lk e^{-x} x^{n+k} \rk.
\end{equation*}
The energy-eigenvalue corresponding to the state (\ref{eq:koyukansu}) is
\begin{equation}
E_{nl}^{\alpha}=\frac{\omega_{\alpha}\hbar}{2}( 3+2l+4n ). 
\label{eq:energy}
\end{equation} 
It should be noticed that we use the unit system of $\hbar=1$ throughout this paper. 

\subsection{Bogoliubov approximation and Fr{\"o}hlich type Hamiltonian} 
In the case of the small number excitation of medium bosons around the impurity 
in comparison with the total condensed boson number $N_0$ 
\cite{Bruderer1,Nakano2},   
we can use the Bogoliubov approximation $b_0\simeq \sqrt{N_0}$, 
where $s=0$ corresponds to the lowest energy level ($n=l=0$). 
With keeping terms in the interaction part
up to the linear order of the excited boson, 
we obtain 
\begin{eqnarray}
\mathcal{H}(\br) &\simeq&
H_{ho}(\br) 
+E^b_0 N_0 
+\sum_{s \neq 0} E_s^b b_s^\dagger b_s
+g N_0 |\phi_0^b(\br)|^2 
\nn 
&& \ +g\sqrt{N_0} \sum_{s\neq0} \ldk \Phi_s(\br)  b_{s} 
+\Phi_s^*(\br) b_{s}^\dagger\rdk,  
\label{FH1y}
\end{eqnarray}
where 
\begin{equation*}
     \Phi_{s=\ltk n,l,m\rtk}(\br)\equiv 
     \sqrt{\frac{1}{4\pi}} R_{00}^{b}(r) R_{nl}^b(r) Y_{lm}( \theta, \varphi). 
\end{equation*}
The Hamiltonian (\ref{FH1y}) can be transformed  
into the same form of the 
Fr{\"o}hlich Hamiltonian of the electron-phonon system, 
and the electron polaron 
was originally studied in \cite{FPZ1} for the polar crystals. 
We will use the Hamiltonian (\ref{FH1y}) in the present paper. 

\section{Cranking of boson states} 
In the present study 
we aim to find the lowest energy state of the Hamiltonian (\ref{FH1y})
for given expectation values of the total angular momentum operators.  
These states correspond to the yrast states 
appeared in the description of rotational collective excitations
of an axially deformed nucleus in nuclear physics,
where the rotation axis is not parallel to that of the axially symmetry, 
and the gauge transformation (cranking) 
$e^{i\omega_k t \hat{J}_k}$ is introduced conveniently   
to shift the state from the normal space-fixed frame 
to the co-rotating frame with the nucleons 
in which the nucleus wave function is stationary \cite{Rowe1,Inglis1,Thouless1}. 
The same method can also be utilized in the present case 
to describe the excitations of bosons around the impurity; 
we rotate the boson cloud around the impurity collectively 
by the gauge transformation ($S$ transformation) 
with the solid angle variables $(\theta, \varphi)$ of the impurity:  
\beq
S(\varphi,\theta) 
&=&
e^{-i\varphi \hat{M}_z} e^{-i\theta \hat{M}_y} 
\eeq
where the boson angular-momentum operator is defined by 
\beq
\hat{M}_i=\sum_{n,l,m,m'}  b_{nlm}^\dagger \lk \hat{\mathcal{L}}_i\rk_{m,m'}^{(l)} b_{nlm'}
\eeq
where $\lk \hat{\mathcal{L}}_i \rk_{m,m'}^{(l)}$ is the matrix element 
of 
a general orbital angular momentum operator $\hat{\mathcal{L}}_i$ 
by the eigen-states of rank $l$ :
\begin{align*}
     & \lk \hat{\mathcal{L}}_i \rk_{m,m'}^{(l)} 
   \equiv  \langle l, m| \hat{\mathcal{L}}_i |l, m'\rangle, \\
     &\quad l =0,1,2, \cdots, \quad
            m,m'=-l,-1+1,\cdots, l-1, l.
\end{align*}
The general form of this transformation has been successfully introduced 
by Schmidt and Lemeshko to investigate the angular momentum distribution 
in the system of a linear rotor impurity 
embedded in bosonic environment in free space \cite{Lemeshko3,Lemeshko4,Lemeshko7}, 
and the simpler version, $S(\varphi,0)$, has been utilized 
in the system of Bose polaron in axially symmetric trap potentials 
for the study of the angular-momentum drag effect \cite{NYI1}. 

\subsection{Cranked angular momentum operators} 
The $S$-transformation practically serves as 
linear transformations for the boson annihilation operators 
and the boson angular-momentum operators:
\beq
S^{-1} b_{nlm} S
&=&
\sum_{m'=-l, \cdots, l}D^l_{m,m'}\lk \varphi, \theta, 0 \rk  b_{nlm'}, 
\label{eq:S}
\\ 
S^{-1} \hat{M}_{i} S
&=&
\sum_{j=0,\pm1} {D^1_{i,j}}^*\lk \varphi, \theta, 0 \rk \hat{M}_{j}, 
\eeq
where we have used the spherical basis: $\hat{M}_0=\hat{M}_z$ and 
$\hat{M}_{\pm1}=\mp \frac{1}{\sqrt{2}} \lk\hat{M}_x\pm i\hat{M}_y\rk$, 
for vector indices, 
and Wigner's $D$ function 
with Euler angles $(\alpha,\beta,\gamma)$ 
for the spacial rotation \cite{Rose1}:
\beq
D^l_{m,m'}\lk \alpha,\beta,\gamma \rk =
\langle l, m| e^{-i\alpha\hat{\mathcal{L}}_z} e^{-i\beta\hat{\mathcal{L}}_y} 
e^{-i\gamma\hat{\mathcal{L}}_z} |l, m'\rangle. 
\eeq
The $S$ transformation acts as a shift operator for the impurity angular-momentum operators:
\begin{eqnarray}
S^{-1}\hat{L}_0S
&=&
\hat{L}_0+S^{-1}\lk \hat{L}_0 S\rk
=
\hat{L}_0-S^{-1}\hat{M}_z S, 
\\
S^{-1}\hat{L}_{\pm1}S 
&=&
\hat{L}_{\pm1}+S^{-1}\lk \hat{L}_{\pm1}S\rk 
\nn 
&=&
\hat{L}_{\pm1} 
+\frac{1}{\sqrt{2}}  e^{\pm i \varphi} 
S^{-1} \ldk 
i\,  e^{-i\varphi \hat{M}_z} \hat{M}_y e^{i\varphi \hat{M}_z} 
\mp \cot\theta  \hat{M}_z
\rdk 
S,  
\end{eqnarray}
where we have used the spherical-basis representation: 
\beq
\hat{L}_0&=&\hat{L}_z=-i\partial_\varphi, 
\\ 
\hat{L}_{\pm1}&=&
\mp \frac{1}{\sqrt{2}} \lk \hat{L}_x\pm i\hat{L}_y\rk
=\frac{1}{\sqrt{2}} e^{\pm i \varphi} \lk
 -\partial_\theta \mp i\, \cot{\theta} \partial_\varphi \rk. 
\eeq 

In the present system, 
the total angular momentum operator of the system is given by
\begin{equation}
\hat{J}_i=\hat{L}_i+\hat{M}_i, 
\end{equation}
and the $z$-th component $\hat{J}_z$ and 
the squared amplitude $\hat{J}^2$ are conserved: 
$\ldk \hat{J}_z, {\mathcal H}(\br) \rdk 
=\ldk \hat{J}^2, {\mathcal H}(\br) \rdk
=\ldk \hat{J}_z,\hat{J}^2\rdk=0$. 
Using the transformation formulas of the angular momentum operators 
$\hat{M}_i$ and $\hat{L}_i$,
the $S$-transformed operators of $\hat{J}_z$ and $\hat{J}^2$ 
becomes,
\beq
S^{-1}\hat{J}_zS
&=&
S^{-1} (\hat{L}_z+\hat{M}_z) S=\hat{L}_z, 
\label{traJz}
\\ 
S^{-1}\hat{J}^2S 
&=& 
S^{-1} \lk \hat{L}^2 +\hat{M}^2+2\hat{M}\cdot \hat{L}\rk S 
\nn  
&=& 
\hat{L}^2+\hat{O}_L +\hat{M}^2
\nn 
&&+2 \sum_{i,j=0,\pm1} \hat{M}_j^\dagger {D^1_{i,j}}(\varphi, \theta, 0)
\ldk 
S^{-1}(\hat{L}_i S)+ \hat{L}_i
\rdk,  
\label{traJsq}
\eeq
where we used the scalar product  
$\hat{M}\cdot \hat{L}=\sum_{i=0,\pm1} \hat{M}_i^\dagger \hat{L}_i$ in the spherical basis representation, 
and the shift operator $\hat{O}_L$ of $\hat{L}^2$ is defined by 
\beq
\hat{O}_L:=S^{-1}\hat{L}^2S-\hat{L}^2=S^{-1}\lk \hat{L}^2 S\rk+ 2S^{-1}(\hat{L} S)\cdot \hat{L}. 
\eeq 

We see from the results (\ref{traJz}) and (\ref{traJsq}) 
that the $z$-component of the total angular momentum is taken over solely 
by the impurity after the $S$ transformation, 
while the total angular momentum of the system looks complicated. 
\footnote{In the case of the linear rotor impurity \cite{Lemeshko3,Lemeshko4}  
the total angular momentum operator is transformed to be that of the impurity, 
which is by virtue of the intrinsic angular momentum of the rotor itself.} 

\subsection{Cranked Hamiltonian} 
In similar calculation, 
the $S$-transformation of the Hamiltonian (\ref{FH1y}) is obtained, 
\begin{eqnarray}
S^{-1} {\mathcal H}(\br) S
&=&
H_{ho}(\br) +\frac{\hat{O}_L}{2m_I r^2}
+ \sum'_{n,l,m}E_{nl}^b 
b^\dagger_{nlm} b_{nlm}
\nn 
&&
+ E^b_{00} N_0 +g N_0 |\phi_0^b(\br)|^2 
\nn 
&& 
+g\frac{\sqrt{N_0}}{4\pi}R_{00}^{b}(r) \sum'_{n,l}  
\sqrt{2l+1}R_{nl}^b(r) \lk b_{nl0}+ b^\dagger_{nl0} \rk, 
\label{traH}
\end{eqnarray}
where the symbols $\sum'_{n,l,m}$ and $\sum'_{n,l}$ represent the summations 
except $n=l=m=0$ and $n=l=0$ respectively. 
In the derivation, 
we have used the formula 
$Y_{lm}^*(\theta,\varphi)=\sqrt{\frac{2l+1}{4\pi}} D_{m,0}^l(\varphi, \theta,0)$ 
and the orthogonality of the $D$ functions \cite{Rose1}. 
The second term including the shift operator $\hat{O}_L$ corresponds 
to the rotation energy of the impurity, 
which comes from the rotation energy of excited bosons originally before the cranking. 
The last term is that of the boson-impurity coupling; 
it should be noticed that it includes the coupling with the excited bosons with $m=0$ 
in the $S$-transformed Hamiltonian\cite{Lemeshko3,Lemeshko4}. 

\section{Variational method} 
Let's develop the variational method 
to obtain the lowest energy states 
under the condition that the azimuthal and $z$ (magnetic) components of the total angular momentum are given by the expectation values $(J, J_z)$. 
The Hamiltonian (\ref{traH}) shows 
that the impurity-boson interaction term includes only the excited bosons with $m=0$ 
after the $S$-transformation, 
so that, as a variational state of excited bosons around the impurity, 
we employ the coherent state for the excited bosons with the quantum numbers $s=( n,l,0 )$:
\cite{LLP1,Schweber1}, 
\beq
\left| b \right\rangle 
&=& \exp
\sum'_{n,l} \lk f_{nl} b_{nl0}^\dagger-f_{nl}^*b_{nl0} \rk  
\left| 0 \right\rangle, 
\label{coh1}
\eeq
where the variational parameters $f_{nl}$ and $f_{nl}^*$ are 
eigenvalues of annihilation and creation operators: 
$b_{nl0} \left| b \right\rangle =f_{nl}\left| b \right\rangle$, 
$\left\langle b \right| b_{nl0}^\dagger=\left\langle b \right| f_{nl}^*$.
The state $\left| b \right\rangle$ is a normalized one: $\langle b|b\rangle=1$. 
It would be a good approximation for the heavy impurity trapped in the deep potential; 
in the case of heavy mass or high trap-frequency limits of the impurity, 
the above coherent state becomes the exact solution
because the impurity becomes localized at the center of trap. 
\footnote{A marginal case where $m_I \rightarrow \infty$ as $m_I\omega_I^2$ is kept finite is also soluble.}

Now we use the abbreviated notation for the expectation value of operator 
by the coherent state $\left| b \right\rangle$ as 
$\langle \cdots \rangle_b \equiv \langle b| \cdots |b\rangle$. 
Then that of the transformed Hamiltonian (\ref{traH}) and 
the $S$-transformed total angular momentum operators become,
\beq
\langle S^{-1} {\mathcal H}(\br) S \rangle_b
&=&
H_{ho}^f(\br) 
+ \sum_{n,l}' \ldk \frac{l(l+1)}{2m_I r^2} + E_{nl}^b \rdk |f_{nl}|^2
\nn 
&&
+ E^b_{00} N_0 +g N_0 |\phi_0^b(\br)|^2
\nn
&&
+g\frac{\sqrt{N_0}}{4\pi}R_{00}^{b}(r) \sum_{n,l}' 
\sqrt{2l+1} R_{n,l}^b(r) \lk  f_{nl} + f^*_{nl} \rk, 
\label{hb1}
\\ 
\langle S^{-1}\hat{J}^2S \rangle_b
&=& \hat{L}^2, 
\label{L2b1}
\\
\langle S^{-1}\hat{J}_z S \rangle_b &=& \hat{L}_z, 
\label{Lzb1}
\eeq
where we have used the expectation value 
$\langle \hat{O}_L \rangle_b =\sum_{n,l}' l(l+1)|f_{nl}|^2$\footnote{
For derivations of the expectation values, 
see Appendix~A.}.

The expectation value (\ref{hb1}), 
where the bosonic degrees of freedom have been eliminated, 
provide the effective Hamiltonian of the impurity, 
and eqs.~(\ref{L2b1}) and (\ref{Lzb1}) are the corresponding effective total angular-momentum operators
represented with the impurity coordinate.
It is very interesting that the latter are the same with the impurity angular momentum;
it gives an essential advantage in the present variational method    
with the condition of the fixed total angular momentum. 

\subsection{Variational states of impurity}
Since the total angular momentum operators
(\ref{L2b1}) and (\ref{Lzb1}) are given 
by those of the impurity, 
the variational state of the impurity can be assumed 
as the eigenfunctions (\ref{eq:koyukansu}) of the impurity 
with fixed azimuthal and magnetic quantum numbers $(J, J_z)$:
\beq
\Psi_{JJ_z}(\br) = \sum_{n} F_{nJJ_z} \, \phi^I_{nJJ_z}(\br) 
\eeq 
where $J_z=-J, -J+1, \cdots, J-1, J$ 
and the coefficients $F_{nJJ_z}$ serve as Ritz-type variational parameters.  
Note that we do not consider mixing of different angular momenta, 
because of rotational symmetry.
Since the states 
with large principal quantum numbers less contribute in the ground state 
in the weak coupling regime, 
we truncate the variational state up to $n=1$ in the present calculation:
\beq
\Psi_{JJ_z}(\br) = \sum_{n=0,1} F_{nJ} \,\phi^I_{nJJ_z}(\br).
\label{ivs1} 
\eeq 
Note that the subscript $J_z$ has been omitted in the variational parameters 
since the rotational symmetry of the system gives the degeneracy for the direction in real space 
and the variational parameters do not depend on $J_z$.
In solving the variational equations,   
we impose the normalization condition for the parameters: $|F_{0J}|^2+|F_{1J}|^2=1$. 

\subsection{Variational energy functional and solutions} 
Now taking the expectation value of the Hamiltonian (\ref{hb1}) 
with respect to the impurity's variational state (\ref{ivs1}), 
we obtain the variational energy functional for the state with the total angular momentum $(J ,J_z)$:
\beq 
E\ldk F_{nJ}; f_{nl}\rdk 
&=& \langle  {\mathcal H}(\br) \rangle_{JJ_z}
\nn
&=& 
E^I_{0J}|F_{0J}|^2+E^I_{1J}|F_{1J}|^2
+ E^b_{00} N_0 \lk |F_{0J}|^2+|F_{1J}|^2 \rk
\nn
&&
+ \sum_{n,l}' \ldk E_{nl}^b \lk |F_{0J}|^2+|F_{1J}|^2 \rk +\frac{l(l+1)}{2m_I}G[F_{0J};F_{1J}]\rdk 
|f_{nl}|^2
\nn 
&&
+g\frac{N_0}{4\pi} H[F_{0J};F_{1J}]_{00}
\nn 
&&
+g\frac{\sqrt{N_0}}{4\pi} \sum_{n,l}'
\sqrt{2l+1}  
\ldk 
H[F_{0J};F_{1J}]_{nl} f_{nl}
+H[F_{0J};F_{1J}]_{nl}^* f^*_{nl} 
\rdk, 
\label{eq:vE}
\eeq 
where we have defined the functionals: 
%
\beq
G[F_{0J};F_{1J}] &=& \int_\br \frac{1}{r^2}
\left|\sum_{n=0,1} F_{nJ} \phi^I_{nJJ_z}(\br)\right|^2, 
\label{functio1}
\\
H[F_{0J};F_{1J}]_{nl} &=& \int_\br {R_{00}^{b}(r)} R_{nl}^b(r)
\left|\sum_{n=0,1} F_{nJ} \phi^I_{nJJ_z}(\br)\right|^2. 
\label{functio2}
\eeq 
The variational equation $\delta E/\delta f_{nl}^*=0$ 
gives the formal solution:
\beq
\bar{f}_{nl}[F_{0J};F_{1J}] =-g\frac{\sqrt{N_0}}{4\pi}
\frac{\sqrt{2l+1} H[F_{0J};F_{1J}]_{nl}^*}
{E_{nl}^b \lk |F_{0J}|^2+|F_{1J}|^2 \rk +\frac{l(l+1)}{2m_I}G[F_{0J};F_{1J}]}, 
\label{fnl1}
\eeq
and, plugging it back to the variational energy (\ref{eq:vE}),   
we obtain
\begin{align*}
E[F_{nJ};\bar{f}_{nl}] 
&=E^I_{0J}|F_{0J}|^2+E^I_{1J}|F_{1J}|^2 \nn 
&+E^b_{00} N_0 \lk |F_{0J}|^2+|F_{1J}|^2 \rk +g \frac{N_0}{4\pi} H[F_{0J};F_{1J}]_{00}
\nonumber\\ 
&-g^2\frac{N_0}{(4\pi)^2} \sum_{n,l}'
(2l+1)  
\frac{|H[F_{0J};F_{1J}]_{nl}|^2}{E_{nl}^b \lk |F_{0J}|^2+|F_{1J}|^2 \rk +\frac{l(l+1)}{2m_I}G[F_{0J};F_{1J}]}. 
\end{align*}
Since the coefficients appearing in the variational energy are all real, 
the solutions of $F_{0J}$ and $F_{1J}$ are also found to be real. 
Using the normalization condition $F_{0J}=\sqrt{1-F_{1,J}^2}$ 
and the analytical expression\footnote{see Appendix~B} of $G[F_{0J};F_{1J}]$, 
we finally obtain 
\begin{align}
E[F_{1J}] &=E^I_{0J}+E^b_{00} N_0 +\lk E^I_{1J}-E^I_{0J}\rk F_{1J}^2 
+E_{\rm bg}[F_{1J}] 
+E_{\rm int}[F_{1J}], 
\label{EF1}
\end{align}
where the background interaction energy is,
\beq
E_{\rm bg}[F_{1J}] 
\equiv g \frac{N_0}{4\pi} 
\ldk H_{00J}^0+ 2F_{1J} \sqrt{1-F_{1J}^2} H_{00J}^c+F_{1J}^2
\lk  H_{00J}^1- H_{00J}^0\rk \rdk, 
\eeq
which comes from the interaction between impurity and background condensed 
bosons corresponding to the term $gN_0|\phi^b_0(\bx)|^2$  in  (\ref{hb1}).  
The interaction energy term $E_{\rm int}[F_{1J}]$ is represented as
\beq
E_{\rm int}[F_{1J}] 
\equiv-\frac{g^2N_0}{(4\pi)^2} \sum_{n,l}'
(2l+1)  
\frac{\ldk H_{nlJ}^0+ 2F_{1J} 
\sqrt{1-F_{1J}^2}H_{nlJ}^c+F_{1J}^2\lk H_{nlJ}^1- H_{nlJ}^0\rk \rdk^2}
{E_{nl}^b +\frac{l(l+1)}{2J+1} \omega_I
\ldk 1+ 2F_{1J} \sqrt{1-F_{1J}^2}/\sqrt{J+\frac{3}{2}}\, \rdk}, 
\label{int1}
\eeq
which is traced back to the parts including $f_{nl}$ in the second and the last terms in (\ref{hb1}), 
and corresponds to the interaction between the impurity and the excited bosons.  
The explicit forms of the coefficients $H_{nlJ}^{0}, H_{nlJ}^{c}, H_{nlJ}^{1}$ 
in $E_{\rm bg}$ and $E_{\rm int}$ are shown also in Appendix~B. 

In experiments, 
the energy shift, 
i.e., the energy with bare impurity and background BEC contributions being subtracted, 
is measurable using the radio-frequency spectroscopy \cite{Jrgensen1,Hu1}; 
in the present theory, 
it is given by the formula:
\beq
\Delta E[F_{1J}] \equiv \lk E^I_{1J}-E^I_{0J}\rk F_{1J}^2+E_{\rm bg}[F_{1J}] +E_{\rm int}[F_{1J}].
\label{shift1}
\eeq

\subsection{Comparison with the second order perturbation theory} 
In general, the solutions of variational method in the present method  
includes non-perturbative effects, 
but it is heuristic and interesting to see its perturbative nature 
before going into numerical results. 
Expanding the variational solution (\ref{EF1}) with the coupling constant $g$,
we obtain 
$F_{1J}=-g\frac{N_0}{4\pi} \frac{H_{00J}^c}{ E^I_{1J}-E^I_{0J}}$, 
to the leading order of $g$;
then the ground state energy becomes 
\begin{eqnarray}
E
&=&
E^I_{0J}+E^b_{00} N_0 +g\frac{N_0}{4\pi}  H_{00J}^0
-g^2\frac{N_0^2}{(4\pi)^2} \frac{\lk  H_{00J}^c\rk^2}{ E^I_{1J}-E^I_{0J}}
\nn 
&&
\quad -g^2\frac{N_0}{(4\pi)^2} \sum_{n,l}'
\frac{(2l+1) \lk H_{nlJ}^0 \rk^2}
{E_{nl}^b +\frac{l(l+1)}{2J+1}\omega_I}. 
\label{pE1}
\end{eqnarray} 
to the order of $g^2$.
The result should be compared with that of the second-order perturbation theory; 
for the ground state energy of $J=0$\footnote{
For derivation, see Appendix~C. }: 
\beq
\langle {\mathcal H}  \rangle 
&\simeq&
E_{00}^I+E_{00}^bN_0+g\frac{N_0}{4\pi}  H_{000}^0
-g^2 \frac{N_0^2}{(4\pi)^2}
\frac{\lk  H_{000}^c \rk^2}{E_{10}^I-E_{00}^I}
\nn
&& 
-g^2 \frac{N_0}{\lk 4\pi\rk^2}
\sum_{n\neq 0}
\lk 
\frac{\lk H_{n00}^c\rk^2}{E_{10}^I-E_{00}^I+E_{n0}^b}
+ \frac{\lk H_{n00}^0\rk^2}{E_{n0}^b}
\rk. 
\label{E2nd}
\eeq
Comparing (\ref{pE1}) with (\ref{E2nd}), 
we find that differences appear in the $g^2N_0$ term,  
which is attributed to the Fr{\"o}hlich-type boson-impurity interaction. 
In the denominator of  (\ref{pE1}),  
the energy of impurity's intermediate states in (\ref{E2nd}) is replaced 
by an averaged rotation energy $\frac{l(l+1)}{2J+1}\omega_I$. 
It can be explained from the cranking transformation and the angular momentum conservation: 
after cranking transformation, all bosons stop to be in rotating states, 
and the impurity rotates instead in order to satisfy the angular momentum conservation;
consequently its effect appears as the rotation energy. 
In the perturbation theory for the ground state, 
the impurity and bosons intermediate virtual states are taken in the order 
from those of low-energy regardless of the angular momentum conservation. 

\subsection{Distribution functions of excited bosons and quasiparticle residue}
Since the quasi-particle properties of the Bose polaron are characterized 
by the virtual boson excitations around the impurity,  
the number of excited bosons around the poralon 
is an important quantity. 
The excited-boson number $N_{nlm}^{JJ_z}$ of bosons with quantum numbers $( n,l,m )$  
for the polaron with the total angular momentum $(J, J_z)$ 
is given by  
\beq 
N_{nlm}^{JJ_z} &=& 
\langle b_{nlm}^\dagger b_{nlm} \rangle_{JJ_z} 
\nn
&=& 
\int {\rm d}\theta \sin\theta {\rm d}\varphi \, |Y_{JJ_z}\lk \theta, \varphi \rk|^2
\left|D^l_{m,0}\lk \varphi, \theta, 0 \rk \right|^2 
\langle b |  b_{nl0}^\dagger b_{nl0} | b \rangle 
\nn
&=&
\left|f_{nl}\right|^2 
\sum_{L=|J-l|}^{J+l} 
\braket{l 0 ; L 0| J 0}^2 \, \braket{J J_z; l m | L J_z+m}^2.
\label{Nb1}
\eeq 
It should be noted that
the dependence on $J_z$ and $m$ in $N_{nlm}^{JJ_z}$ comes 
through the Clebsch-Gordan coefficients\footnote{
For the definition of the Clebsch-Gordan coefficients, see \cite{Rose1,Edmonds1}.} 
$\braket{l 0 ; L 0| J 0}$ and $\braket{J J_z; l m | L J_z+m}$ 
which are originated 
in the averaged overlap of the coupled angular-momentum states from $J$ and $l$. 
The dependence through the Clebsch-Gordan coefficients is not dynamical but kinematical; 
it can be understood from 
independence of the polaron energy-functional on $J_z$ or $m$. 

From (\ref{Nb1}), 
we obtain the total excited-boson number by summing up quantum numbers:
\beq
N_b ^{J}&=&\sum'_{n,l}\sum_{m=-l, \cdots, l} N_{nlm}^{JJ_z}
=\sum'_{n,l}\left|f_{nl}\right|^2.
\label{nbtot1}
\eeq
It is clear that the total number does not depend on $J_z$ 
but it has the implicit $J$-dependence through the variational parameter $F_{1J}$. 

The real-space density distribution of the excited bosons is given by
the expectation value:
\beq
\left\langle \phi^{b\, \dagger}(\bx) \phi^b(\bx) \right\rangle_{JJ_z} 
&=& 
\sum'_{n,l}\sum_{m=-l, \cdots, l} 
N_{nlm}^{JJ_z}\, |\phi^{b}_{nlm}(\bx)|^2. 
\label{nbx1}
\eeq

For the angular momentum, 
the boson contribution is found to vanish,
\beq
\langle \hat{M}_i\rangle_{JJ_z}=0, 
\eeq
which implies that the impurity alone bears the contribution for $J_z$;
it shows that no drag effects exist for the angular momentum 
unlike the axial symmetric case \cite{NYI1}. 
There are two reasons for this property.
First, thanks to the complete rotational symmetry 
the energy functional becomes spherically symmetric 
and does not depend on $J_z$.  
Second, no angular-momentum exchange can happen between impurity and bosons
through the impurity-boson interaction 
because a density-density type interaction is employed in this work. 
In order for $\langle \hat{M}_i \rangle$ to be finite, 
an asymmetry with respect to $m$ 
is required
in the distribution function of excited bosons, 
but there is no sources of the asymmetry in the present case 
because of the rotational symmetry. 
In the case of axial symmetric trap potentials, 
this specific axis provides an asymmetry for the energy functional and the distribution function \cite{NYI1}. 
We will come back to this point when we present the numerical results in the next section. 

The quasi-particle residue is defined as
\beq
Z_{J}
&=&
\left| \int_\br \phi_{0JJ_z}^{I*}(\br)\langle 0|S|b\rangle \Psi_{JJ_z}(\br) \right|^2
=|F_{0J}|^2 e^{-\sum_{n,l}'|f_{nl}|^2}. 
\label{res1}
\eeq
It also quantifies the modification of the impurity 
due to the interaction effects, 
which is given by the overlap between the bare and interacting impurity states 
with the angular momentum $(J, J_z)$.  
Eq.~(\ref{res1}) shows that 
the residue is factorized 
into the ground state component of the impurity wave function $|F_{0J}|^2$
and a weight factor $e^{-\sum_{n,l}'|f_{nl}|^2}$ of the excited bosons,
while in the spatially uniform case it depends only on the latter. 

\section{Numerical results and discussion}
In numerical calculation 
we take K${}^{40}$ as the impurity immersed in medium bosons of Rb${}^{87}$; 
the trap frequencies of the impurity and the medium bosons  
and the condensed-boson number that we take are
\beq
&& \omega_I=200\, {\rm Hz}, \quad \omega_b=100\,  {\rm Hz}, \quad  N_0 = 10^4, 
\label{ref1}
\eeq 
throughout numerical calculations.
We treat the boson-impurity scattering length as a variable parameter,  
but neglect the boson-boson scattering length in the present calculation.
In actual experiments of Bose polarons \cite{Jrgensen1,Hu1}, 
the trap potentials for impurity and medium-bosons are both axially symmetric, 
and the boson-boson scattering length is usually set to be a small positive number 
to stabilize the boson sector.
In the present theoretical study of the idealized system, 
the zero-point energy in the trap supports and stabilizes the system, 
and the present system of the negligible boson-boson scattering length 
can be potentially realized in real experiments. 

The average density of condensed bosons in the trap system is defined as
\beq
\bar{n} &=&N_0 \int_\br |\phi_0^b(\br)|^4
=N_0\lk \frac{m_b\omega_b}{2\pi}\rk^{\frac{3}{2}}.
\eeq 
We also introduce the scale factors for momentum and energy: 
\beq
k_{\rm ref}=\lk 6\pi^2\bar{n}\rk^{1/3}, 
\quad
E_{\rm ref}=\frac{k_{\rm ref}^2}{2m_b}, 
\eeq
as in the case of the uniform systems \cite{Hu1,Rath1,Li1}.  

\subsection{The ground state energies for the states of small angular momentum} 
\begin{figure}[H]
  \begin{center}
    \begin{tabular}{cc}
 \resizebox{80mm}{!}{\includegraphics{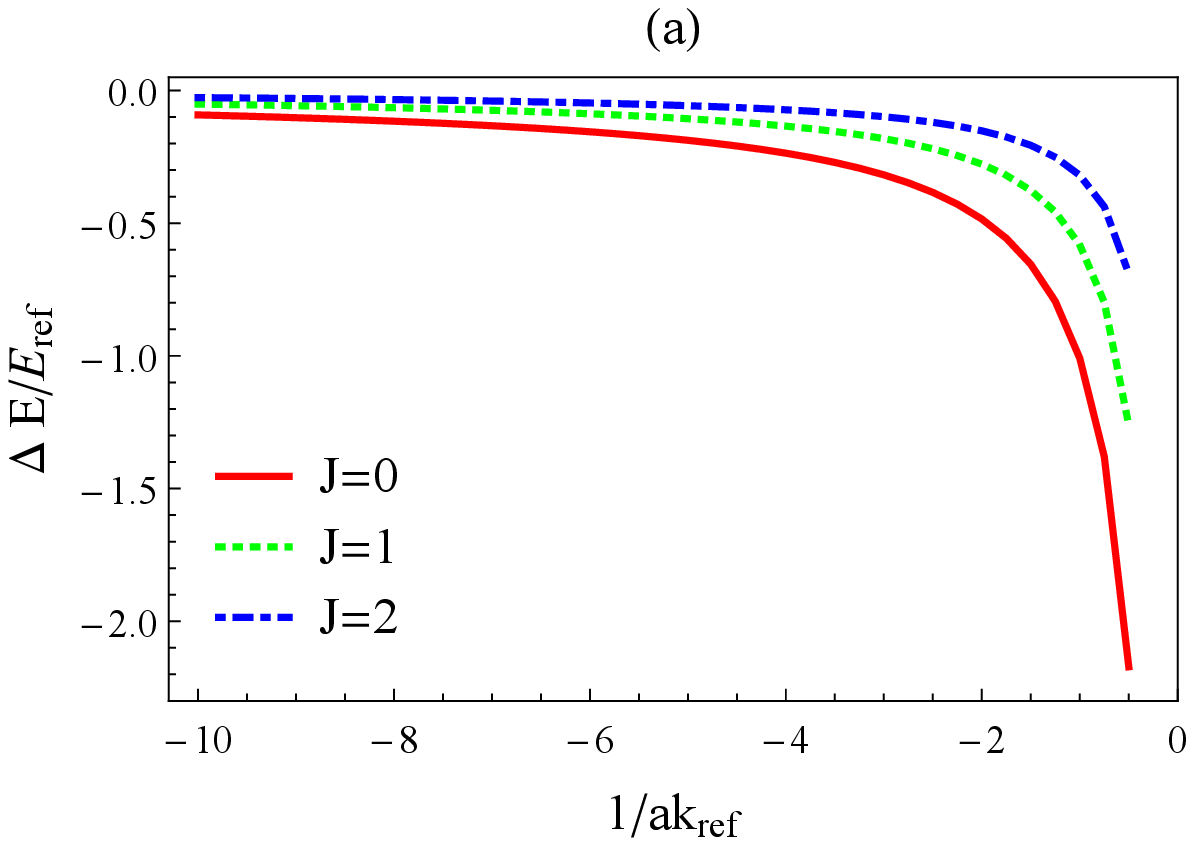}} & 
 \resizebox{80mm}{!}{\includegraphics{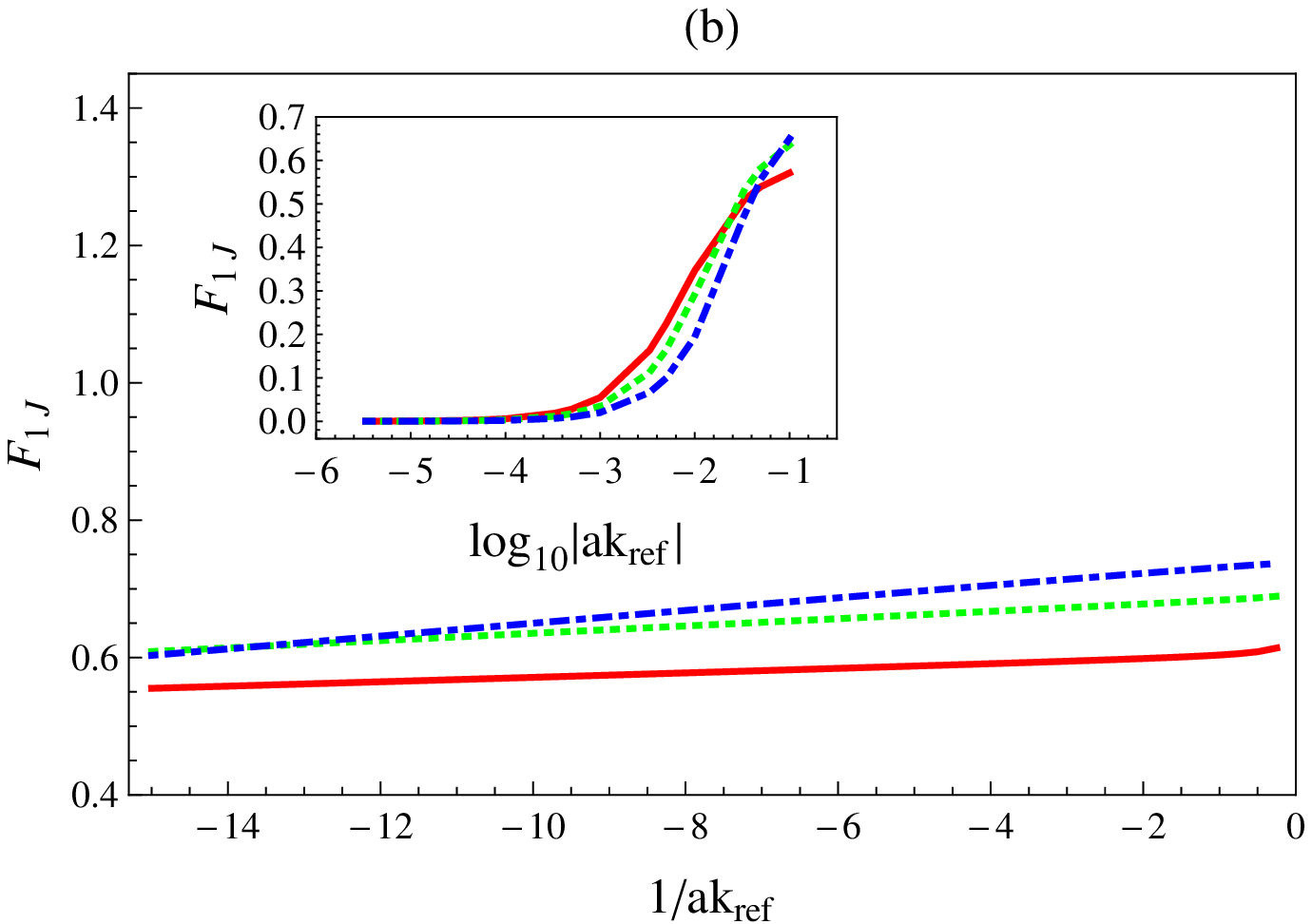}} \\
 \resizebox{80mm}{!}{\includegraphics{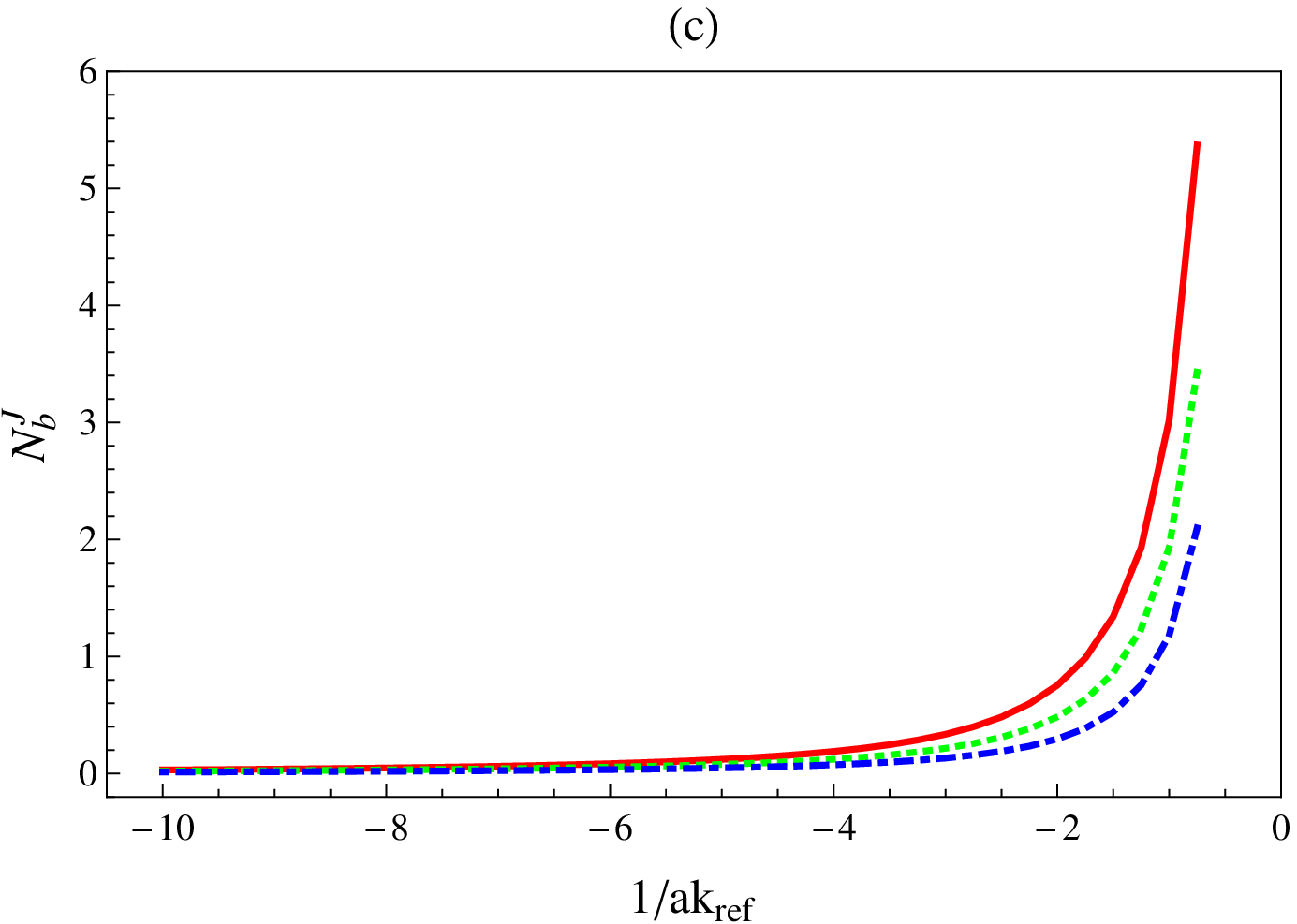}} & 
 \resizebox{80mm}{!}{\includegraphics{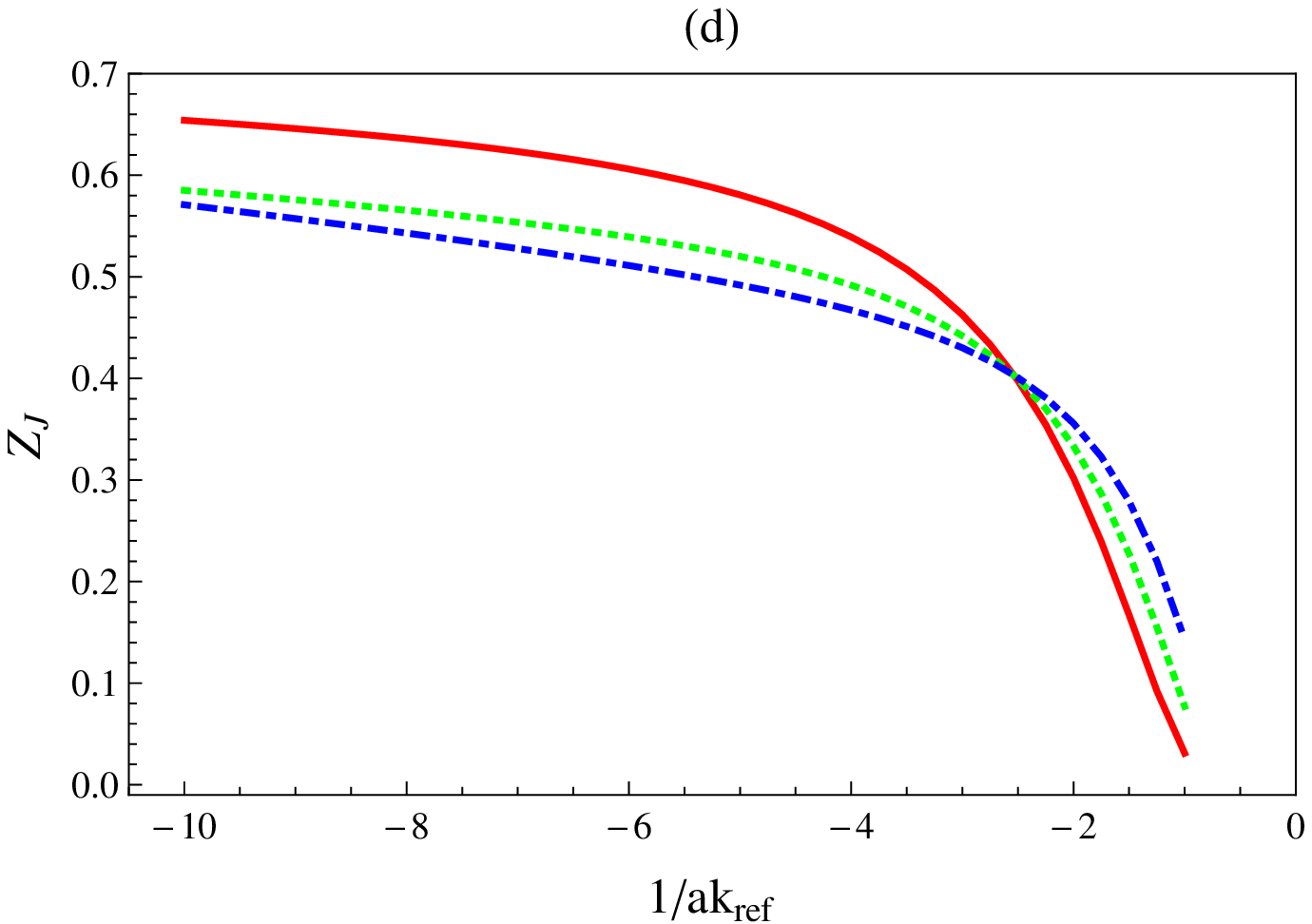}} \\ 
    \end{tabular}
 \caption{The energy shift (a), 
 the variational parameter $F_{1J}$ (b), 
 the total excited-boson number (c), and
 the quasi-particle residue (d) for $J=0,1,2$,
 as functions of the inverse of scattering length
 with the parameters in (\ref{ref1}).
The inset in the panel (b) is for smaller scattering-length region. 
The definitions of these quantities are given in 
  (\ref{shift1}),  (\ref{ivs1}),  (\ref{nbtot1}), 
and  (\ref{res1}), respectively.}
    \label{fig1}
  \end{center}
\end{figure}

In Fig.~\ref{fig1}, 
we show the scattering-length dependence of the ground state properties of polaron:  
the energy shifts (\ref{shift1}), 
the calculated values of impurity's variational parameter $F_{1J}$ in (\ref{ivs1}), 
the total number of excited bosons (\ref{nbtot1}), 
and quasi-particle residue (\ref{res1}), 
for the small numbers of the total angular momenta ($J=0,1,2$). 
\footnote{ 
In these calculations we have taken the approximation 
to cut the summation in the interaction energy (\ref{int1}) up to $(n,l)=(30,10)$. 
We have checked the approximation numerically 
by raising the maximum values of $n$ and $l$ by $100\, \%$;
then the numerical results change within a few $\%$, 
and the sum of $l$ shows a rapid convergence. 
Also, since $H_{nlJ}^{0,c,1} \rightarrow 0$ as $n\rightarrow \infty$ or $l\rightarrow \infty$, 
the series of $n,l$ summation in (\ref{int1}) drop faster than the order of $1/n$ ($1/l$) 
for large number of $n$ ($l$), 
which implies the series is a convergent one. 
} 
%
The energy shift obtained here should be comparable 
with the experimental result \cite{Hu1} 
only for the case of small scattering lengths, 
roughly of $1/ak_{\rm ref} <-2$; 
it is because 
the Bogoliubov approximation (\ref{FH1y}) employed here works 
only if the number of excited bosons is less than or equal to the number of impurities, 
(it is the unity in the present calculation), 
and also the two-level approximation 
in the impurity wave function (\ref{ivs1}) is valid 
for the smaller values of variational solutions  ($F_{1J}\ll F_{0J}$), 
and loses the validity when $F_{1J} \ge 1/\sqrt{2}\sim 0.7$. 
\footnote{ 
Note that the variational solution of $F_{1J}$ is determined 
mainly from the first two terms in (\ref{shift1}), 
and takes smaller values 
in the cases of  the heavier impurity masses or 
of the larger trap frequencies than the present ones. 
}
Also, the behavior of the residue implies that quasi-particle picture of the polaron 
works for about $1/ak_{\rm ref} <-2$ as well. 
In the case of the strong coupling regime and around the unitary limit, 
i.e., $|1/ak_{\rm ref}|<1$, 
we need to include effects of the two-to-two scattering processes 
between impurity and excited boson
which were discarded in the Bogoliubov approximation; 
they are responsible for the effective in-medium shift 
in the unitary limit \cite{Shchadilova2} 
and the in-medium few-body bound states \cite{Levinsen2,Shchadilova2}. 

\subsection{Distributions of excited bosons}
\begin{figure}[H]
  \begin{center}
    \begin{tabular}{c}
 \resizebox{90mm}{!}{\includegraphics{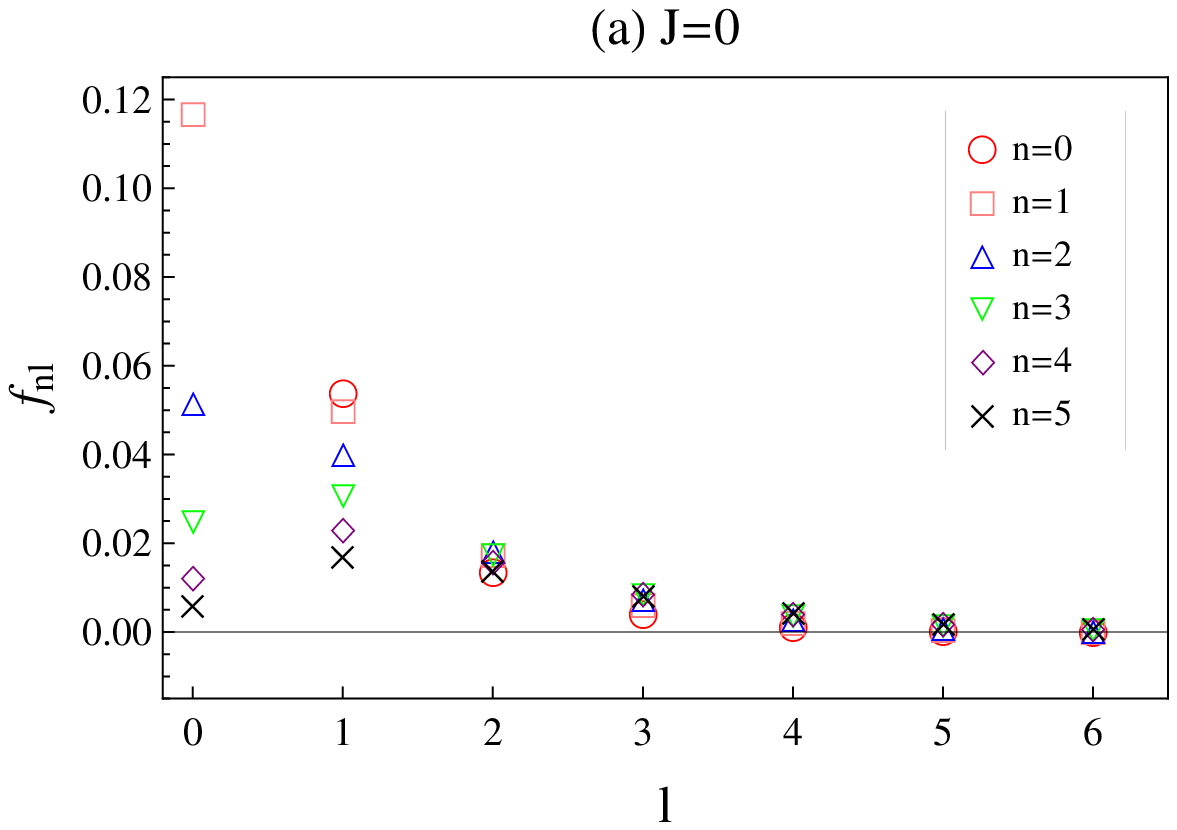}} \\
 \resizebox{90mm}{!}{\includegraphics{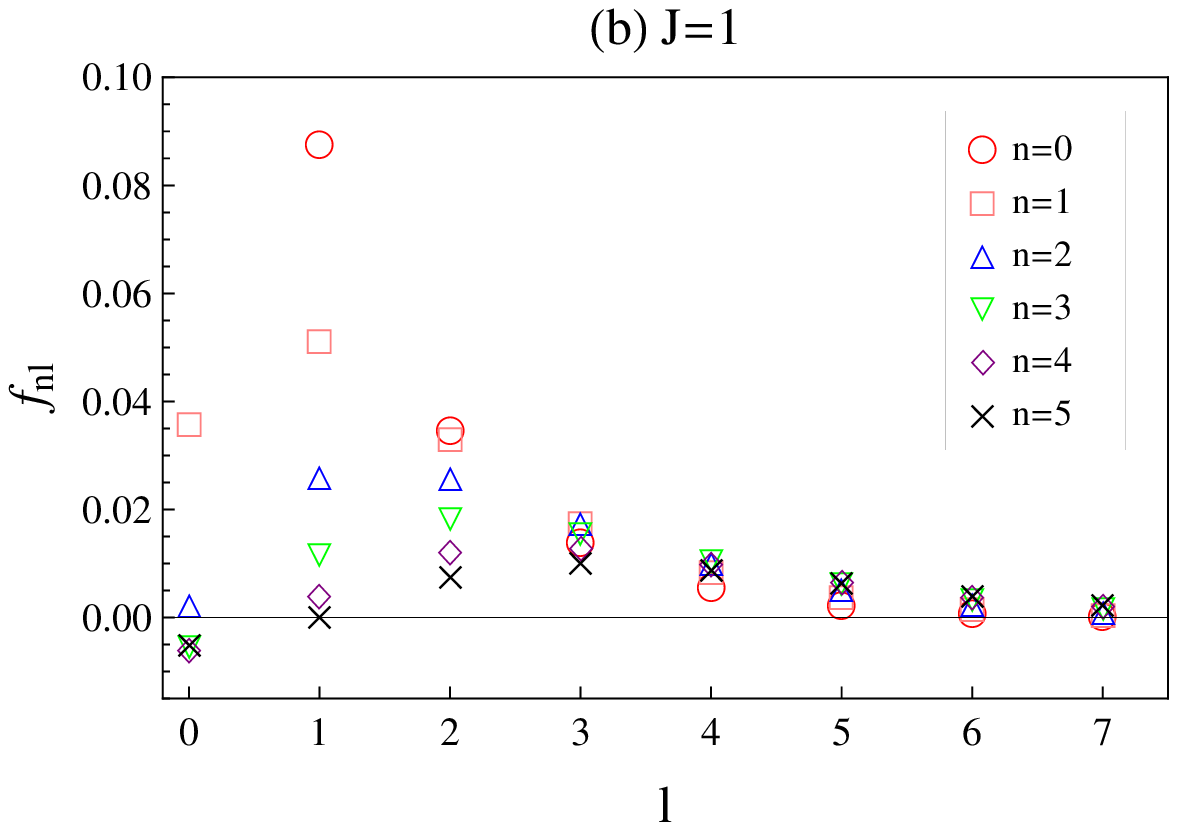}} \\
 \resizebox{90mm}{!}{\includegraphics{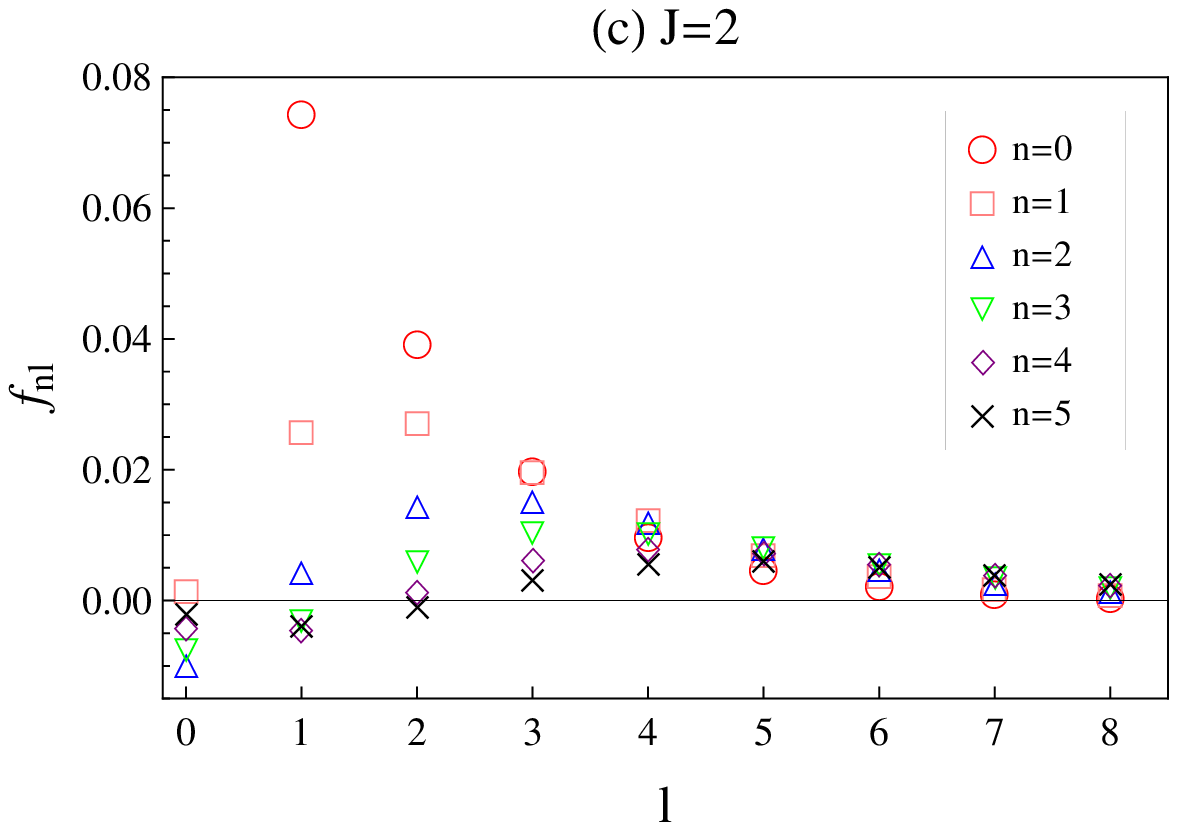}} 
    \end{tabular}
 \caption{The variational parameter of excited bosons (\ref{fnl1}) for $J=0,1,2$ 
and their angular-momentum $l$-dependences. 
A set of parameters is given in (\ref{ref1}), and $1/ak_{\rm ref}= -9.95$. }
    \label{fig2}
  \end{center}
\end{figure}

In Fig.~\ref{fig2} we show the solutions of variational parameter $f_{nl}$ 
for $J=0,1,2$, 
where we have set the impurity-boson scattering length by $a=-5.77 {\rm nm}$ 
corresponding at $1/ak_{\rm ref}= -9.95$. 
The parameter $f_{nl}$ can be interpreted as the probability amplitude 
of excited bosons with the quantum numbers $(n,l)$.
%
%
These figures implies that, 
for each principal quantum number $n$, 
the peak positions of $f_{nl}$ for the quantum number $l$ move to the right 
as the total angular momentum $J$ is increased. 
This is due to the attractive density-density-type interaction 
between impurity and bosons, 
which cause the large overlap between their wave functions 
to lower the interaction energy.   
It can be shown more directly 
in the real space distributions (Fig.~\ref{fig4}-\ref{fig6}). 

\begin{figure}[H]
  \begin{center}
    \begin{tabular}{cc}
 \resizebox{90mm}{!}{\includegraphics{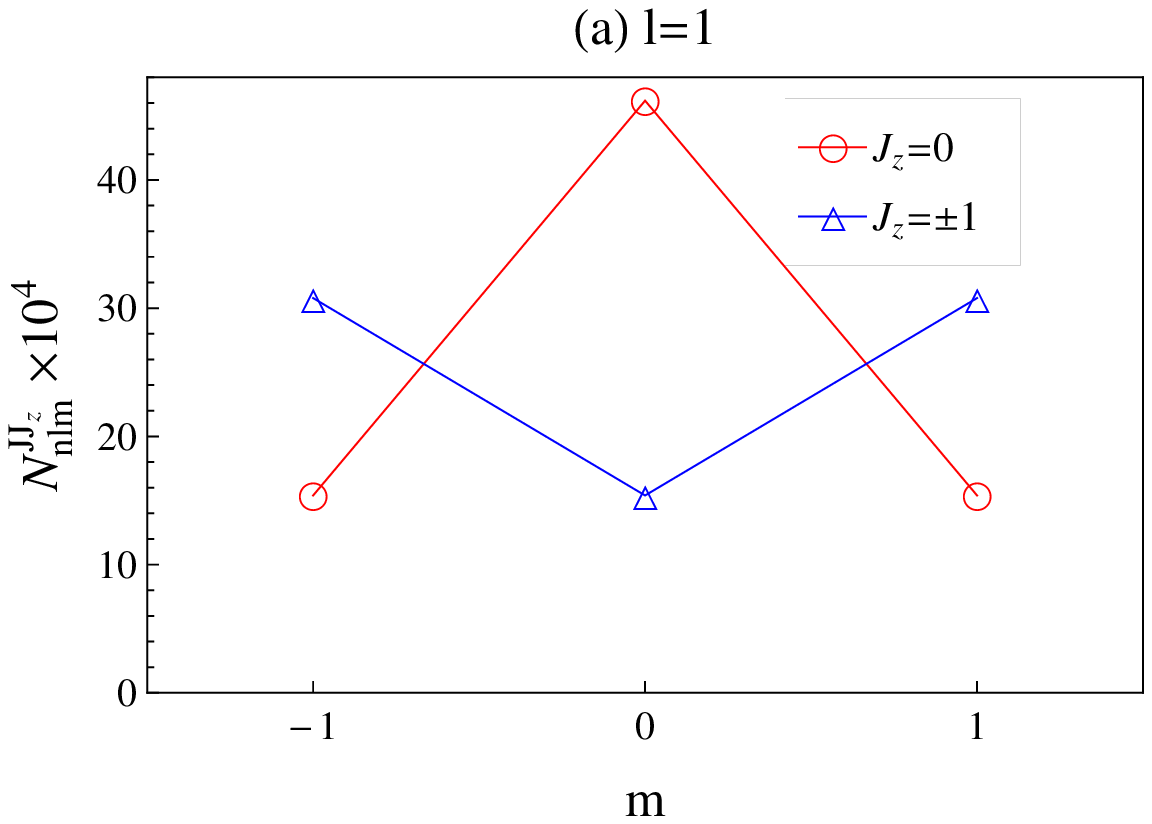}} &  
\resizebox{90mm}{!}{\includegraphics{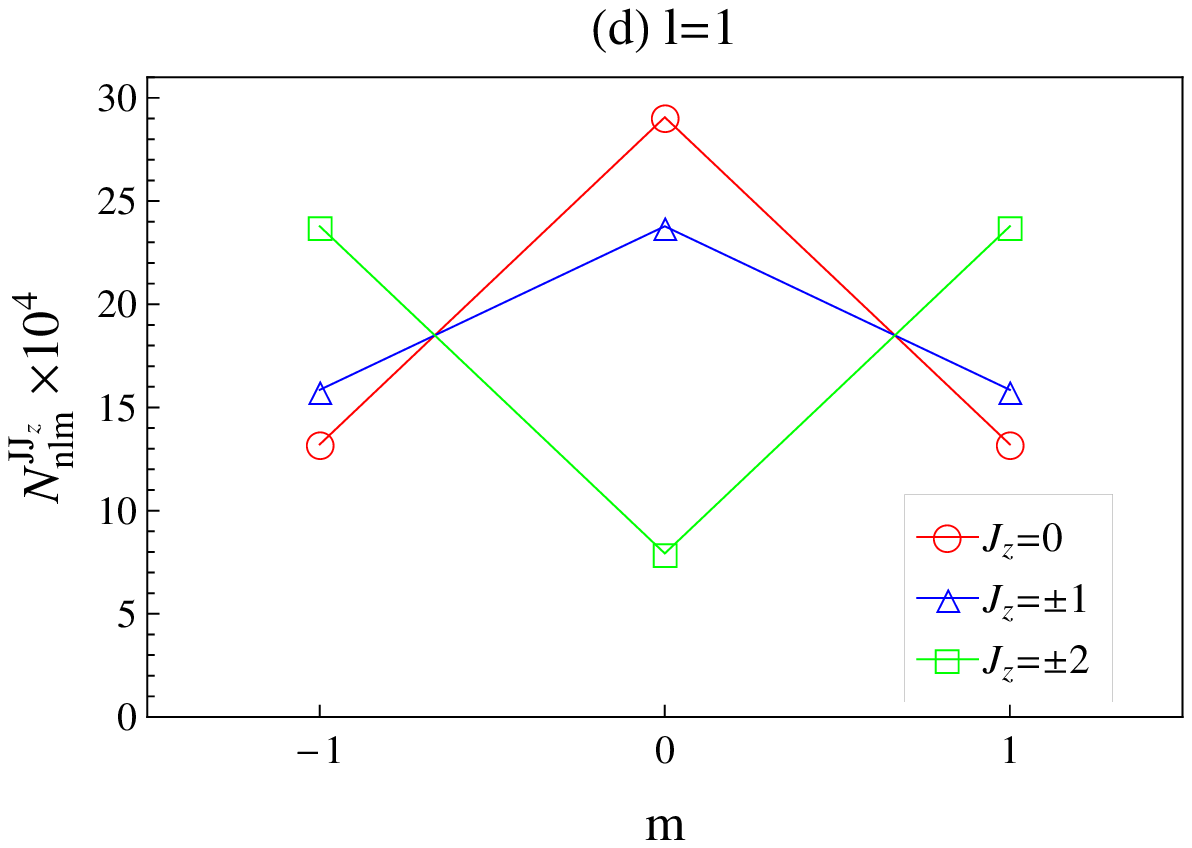}} \\
 \resizebox{90mm}{!}{\includegraphics{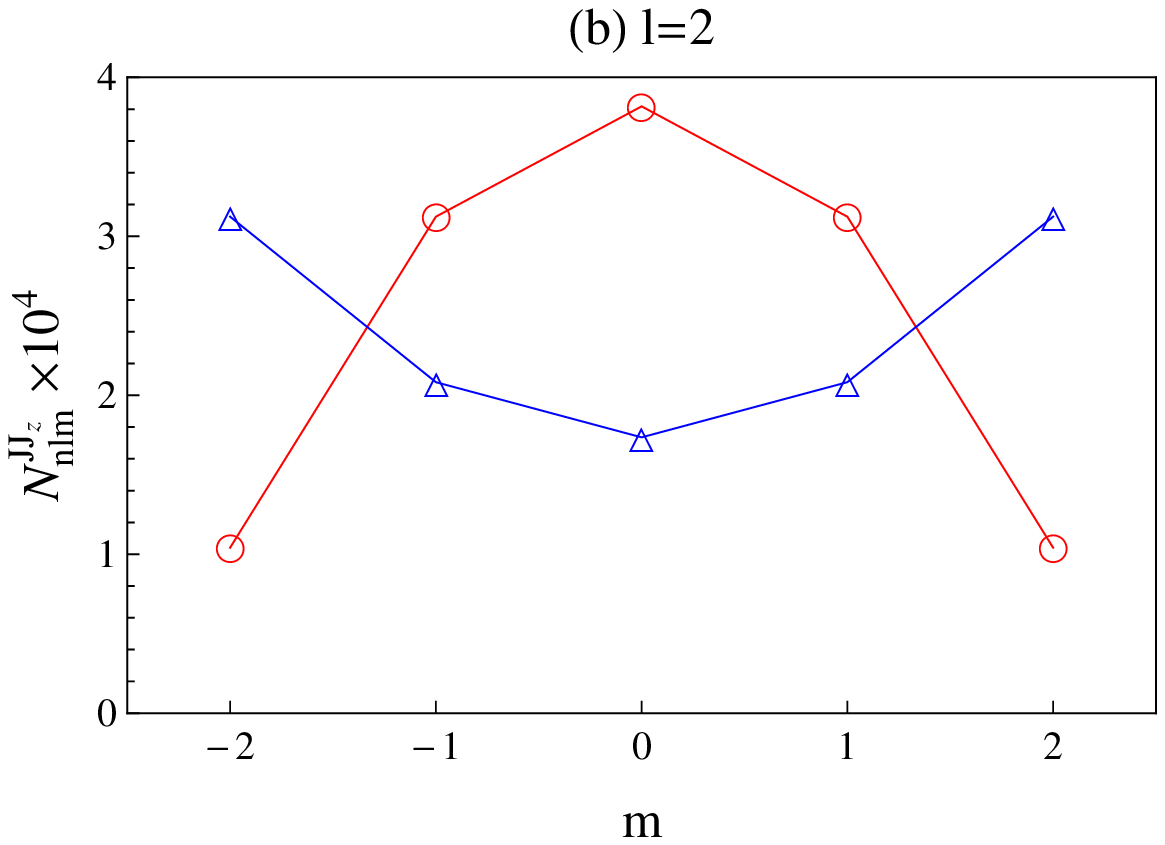}} &
 \resizebox{90mm}{!}{\includegraphics{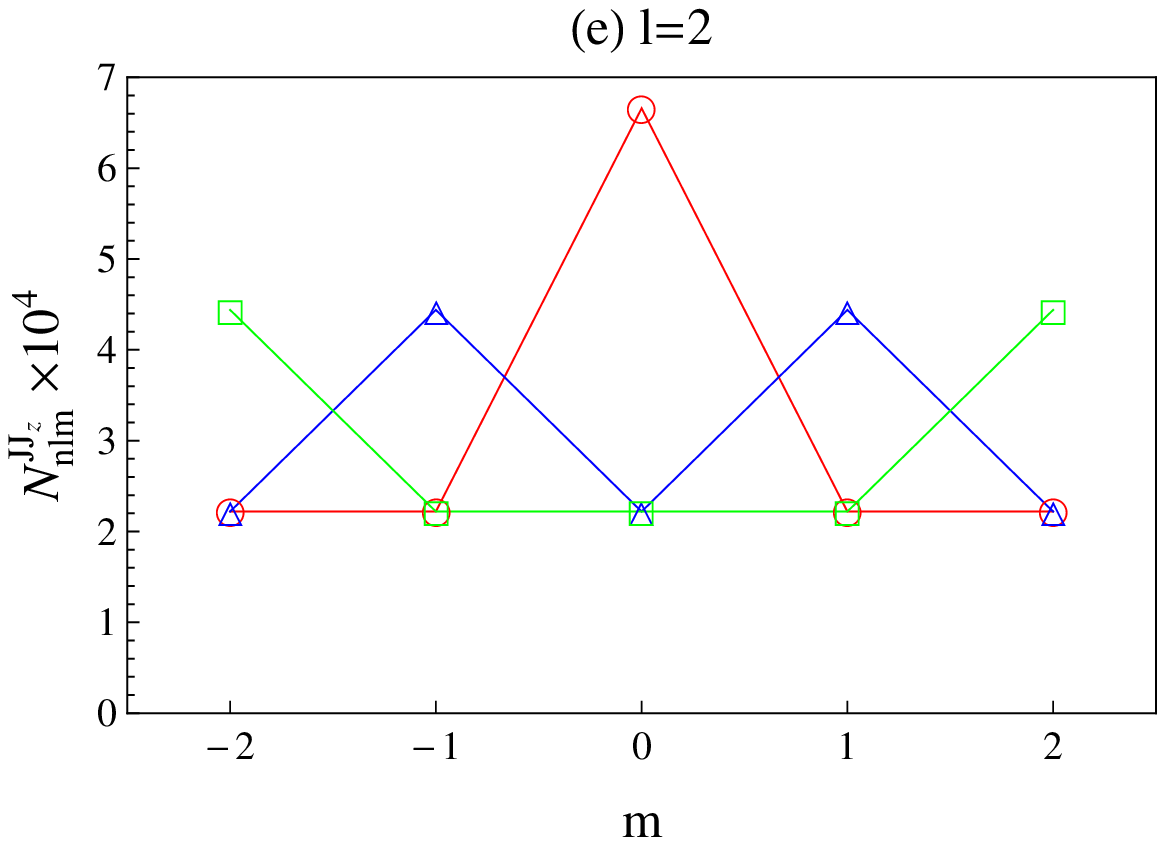}} \\
 \resizebox{90mm}{!}{\includegraphics{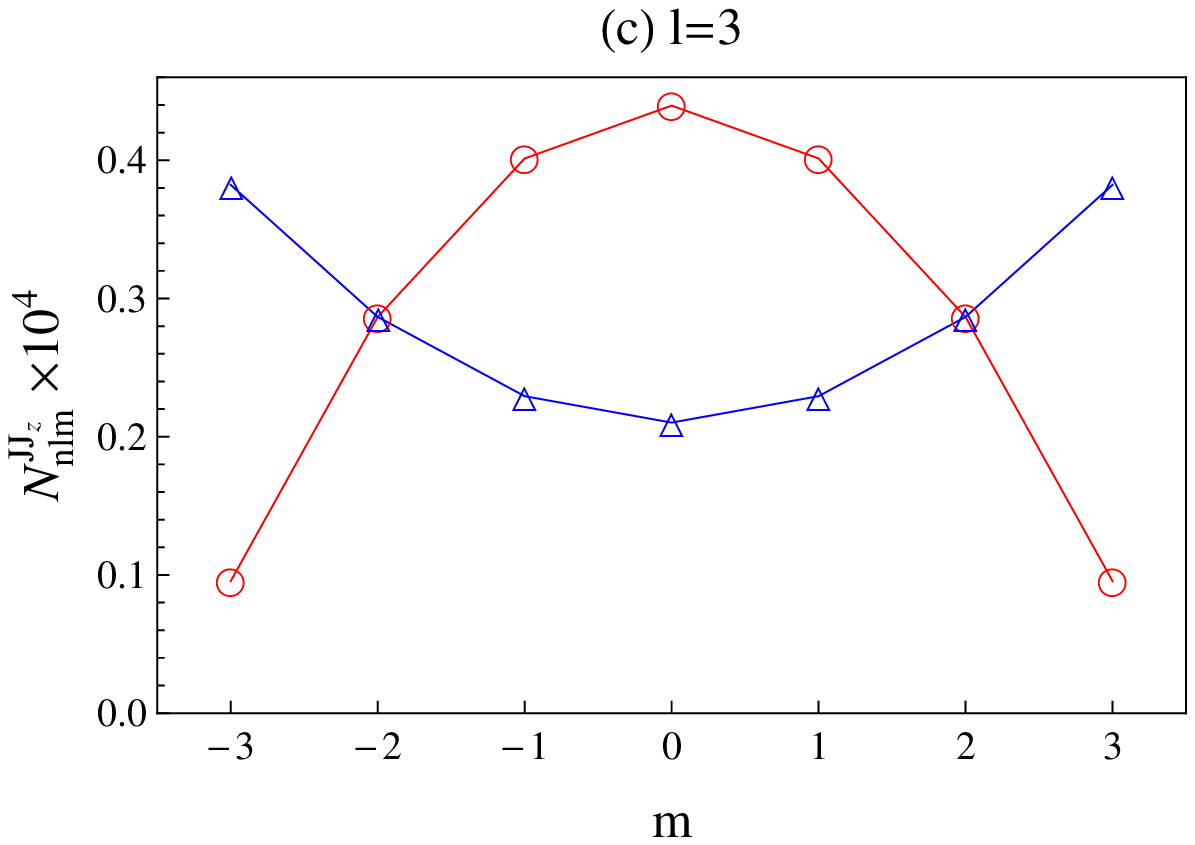}} &
 \resizebox{90mm}{!}{\includegraphics{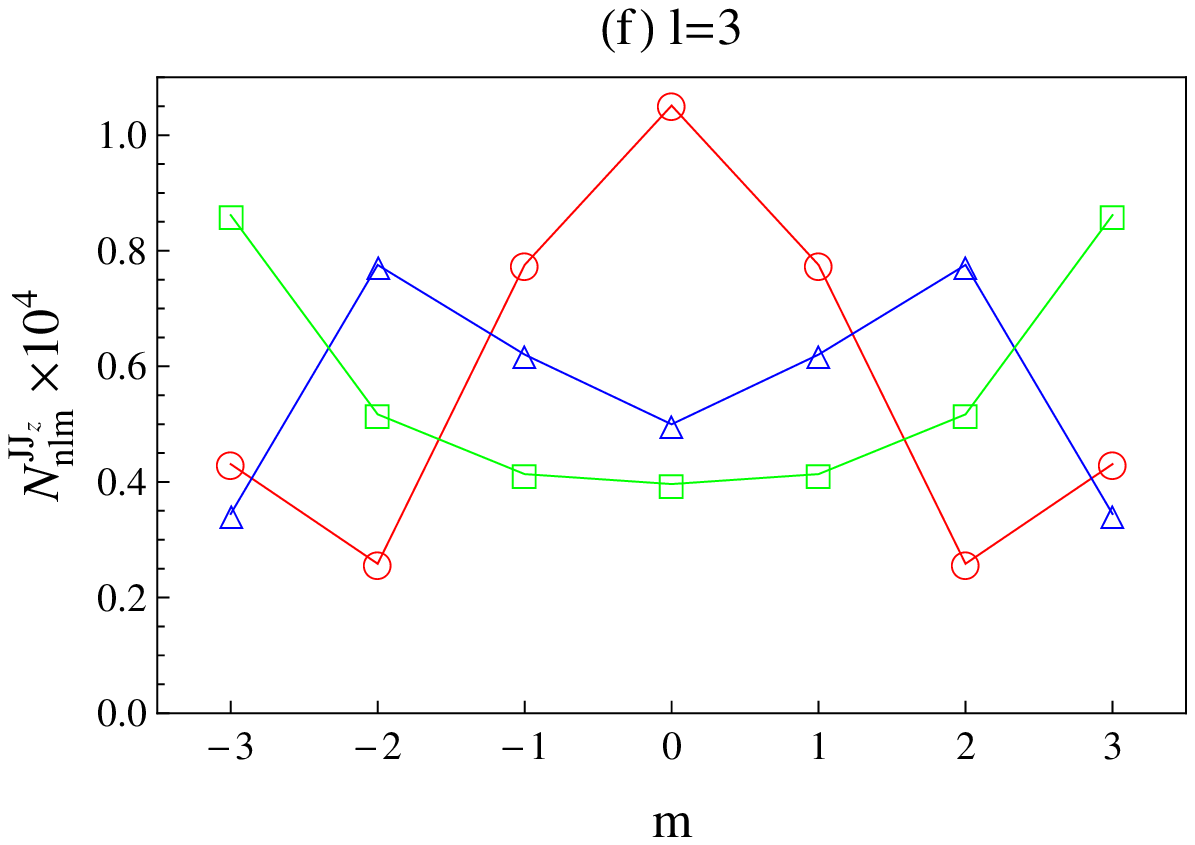}}  
    \end{tabular}
 \caption{
The angular-momentum $(l,m)$-dependences of the excited-boson numbers 
in (\ref{Nb1})  for $n=0$ and $l=1,2,3$. 
Panels (a), (b), (c) in left columns  
and (d), (e), (f) in  right columns 
are for $J=1$ ($J_z=0,\pm1$) and 
$J=2$ ($J_z=0,\pm1,\pm2$), respectively.
The parameter set is the same as in Fig.~2. }
    \label{fig3}
  \end{center}
\end{figure}

\begin{figure}[H]
  \begin{center}
    \begin{tabular}{cc}
 \resizebox{70mm}{!}{\includegraphics{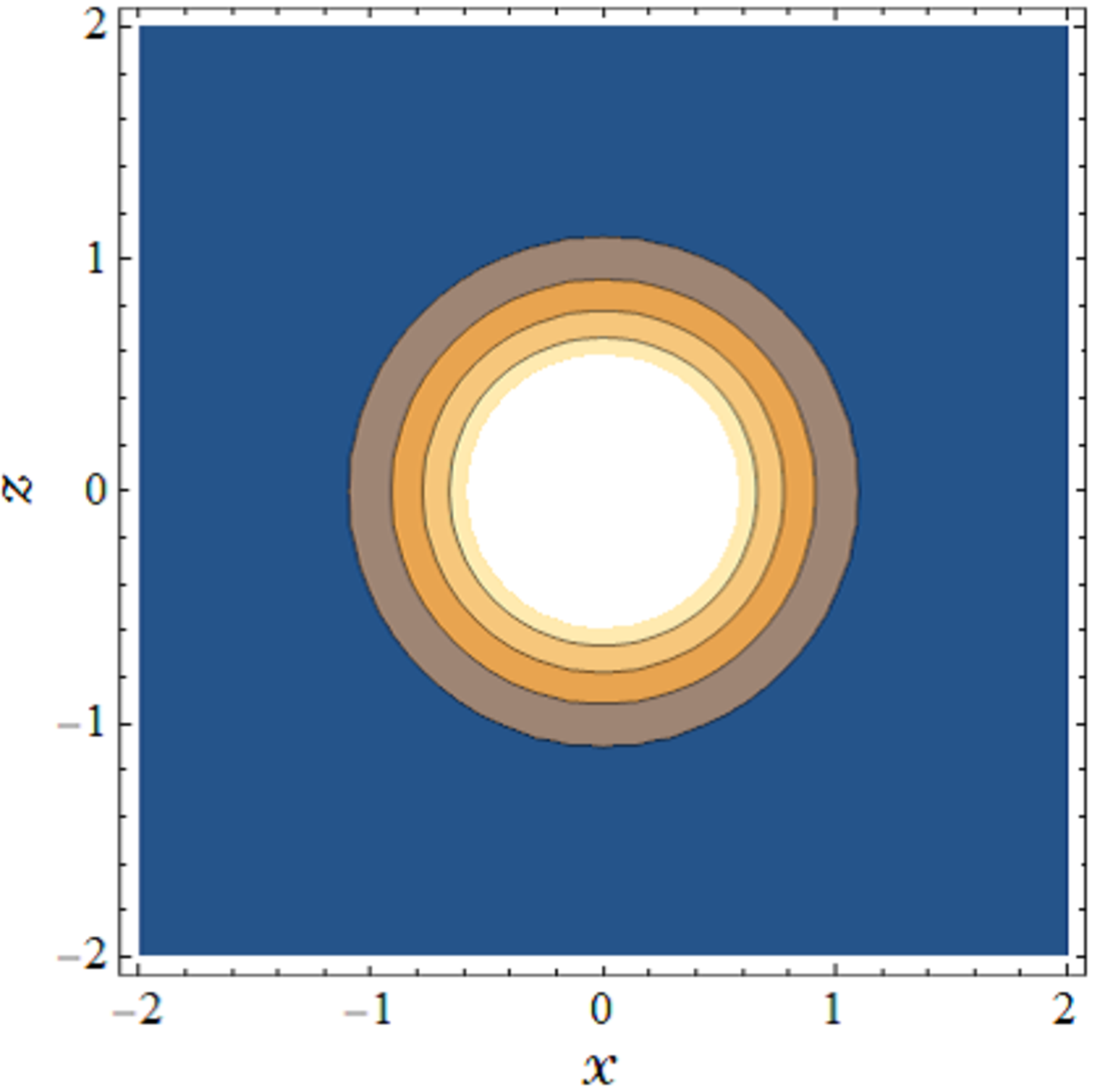}} & 
 \resizebox{70mm}{!}{\includegraphics{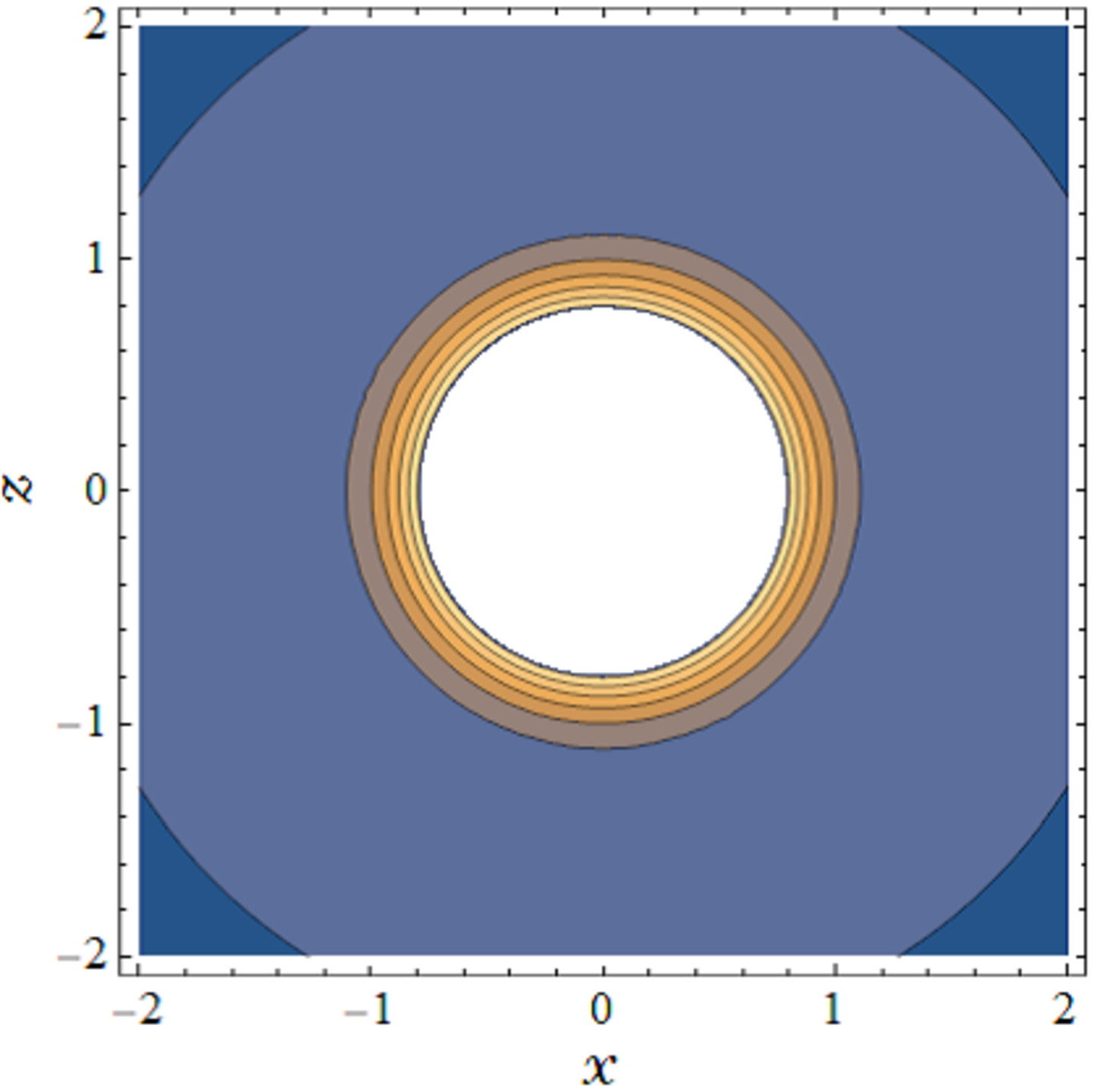}} 
    \end{tabular}
 \caption{ The $J=J_z =0$ contour plots of the impurity's probability density 
 $|\Psi_{JJ_Z}(\br)|^2$ (left panel) 
 and the real-space excited-boson distributions (right panel) defined 
 in (\ref{nbx1}), 
in the cross-section plane of $y=0$, 
where the ordinate is the $z$ axis (the direction of the magnetic quantum number),
 and the abscissa is the $x$ axis.
Note that these plots have the rotational symmetry around the $z$ axis. 
The units of axes are $\lk \omega_I m_I\rk^{-1/2}$ (left) 
and $\lk \omega_b m_b\rk^{-1/2}$ (right), 
and the heights of the contour lines (not shown explicitly) are normalized 
by $\lk \omega_I m_I\rk^{3/2}$ (left) and $\lk \omega_b m_b\rk^{3/2}$ (right). 
The parameter set is the same as in Fig.~2.
}
    \label{fig4}
  \end{center}
\end{figure}

\begin{figure}[H]
  \begin{center}
    \begin{tabular}{cc}
 \resizebox{70mm}{!}{\includegraphics{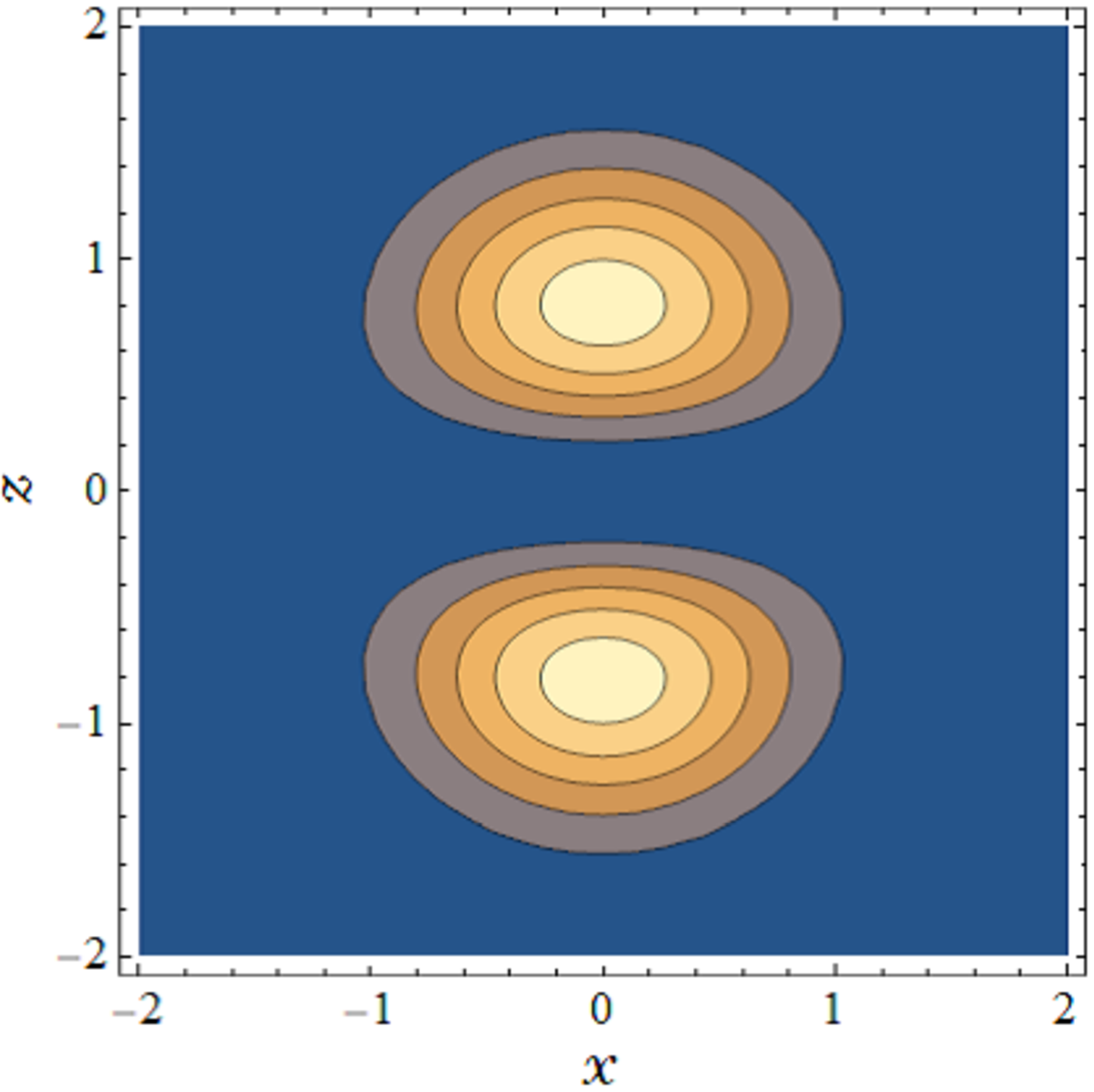}} & 
 \resizebox{70mm}{!}{\includegraphics{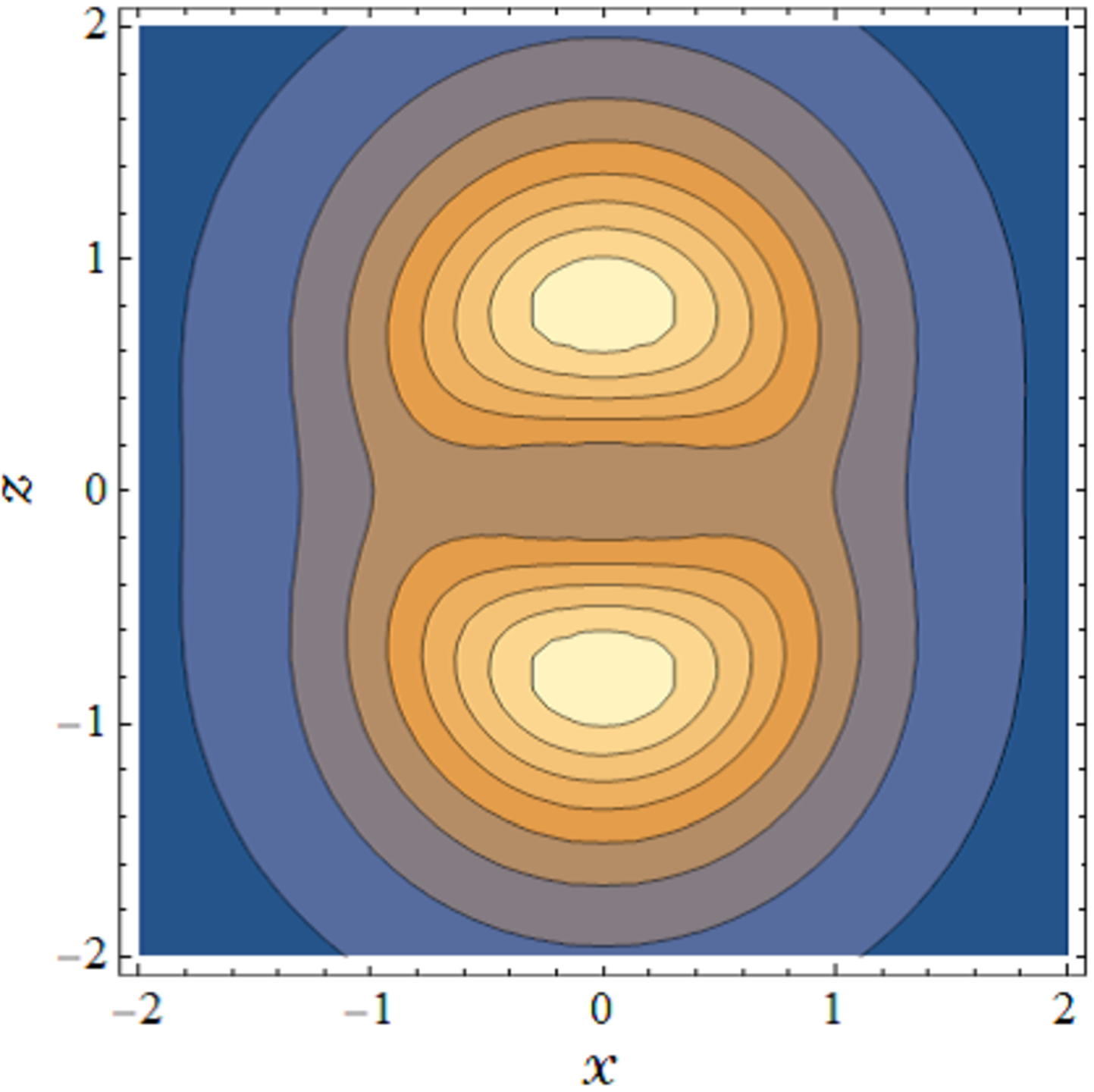}} \\
 \resizebox{70mm}{!}{\includegraphics{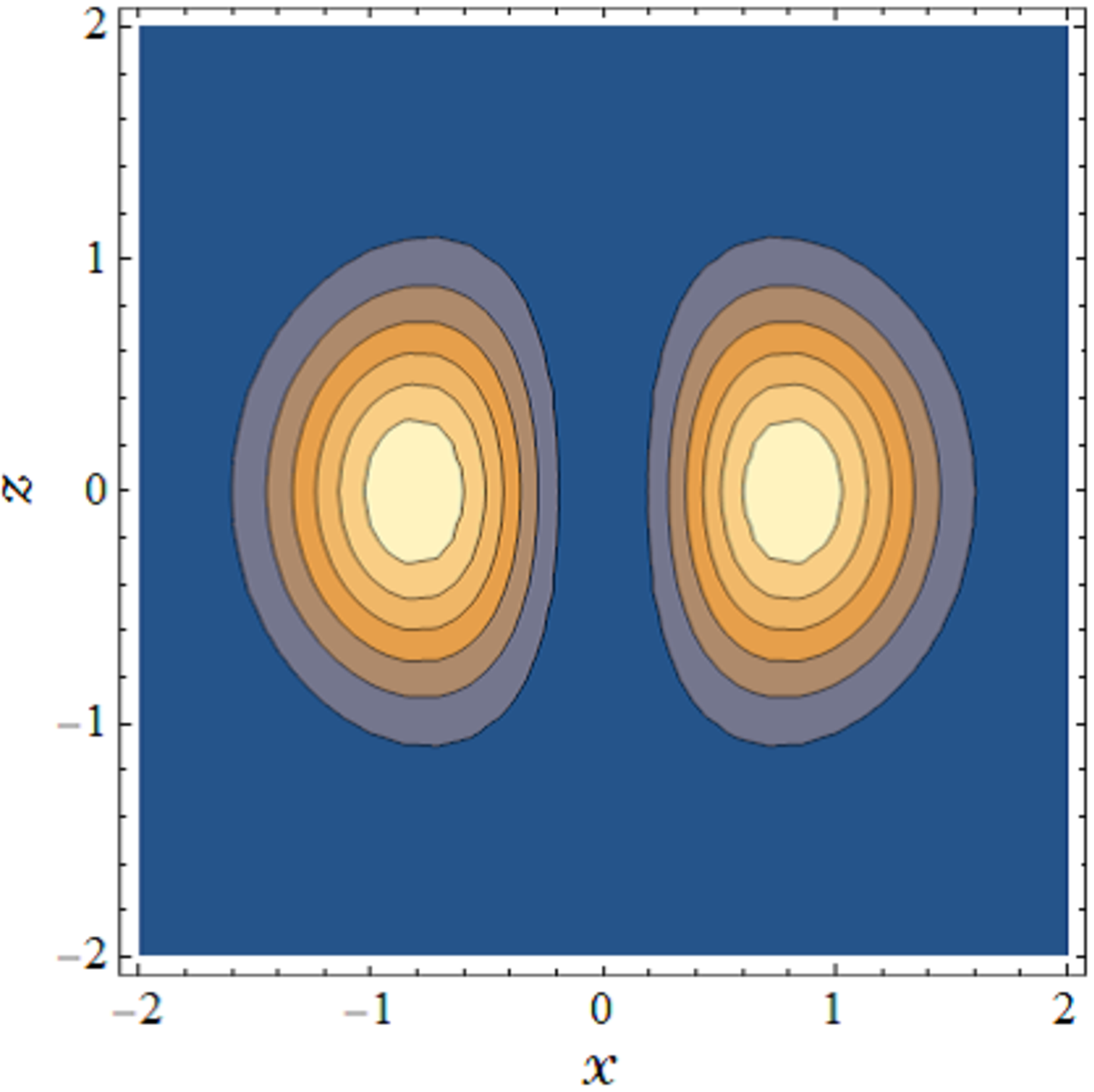}} & 
 \resizebox{70mm}{!}{\includegraphics{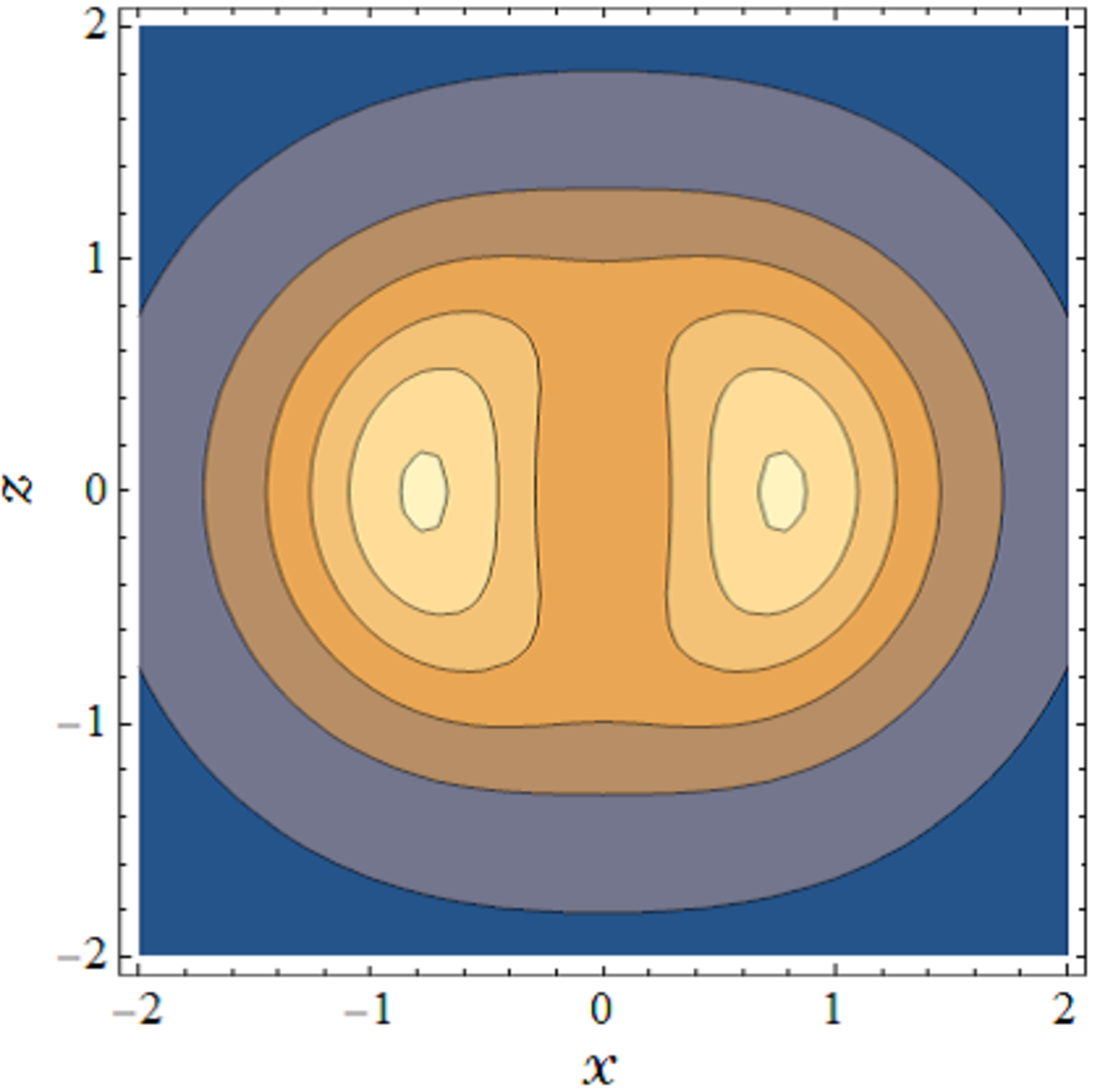}} 
    \end{tabular}
 \caption{The $J=1$ contour plots of the impurity's probability density (left panel) 
 and the real-space excited-boson distributions (right panel):
 the top and bottom panels are for $|J_z|=0, 1$. 
For other explanations, see the caption in Fig.~\ref{fig4}.  }
    \label{fig5}
  \end{center}
\end{figure}

\begin{figure}[H]
  \begin{center}
    \begin{tabular}{cc}
 \resizebox{70mm}{!}{\includegraphics{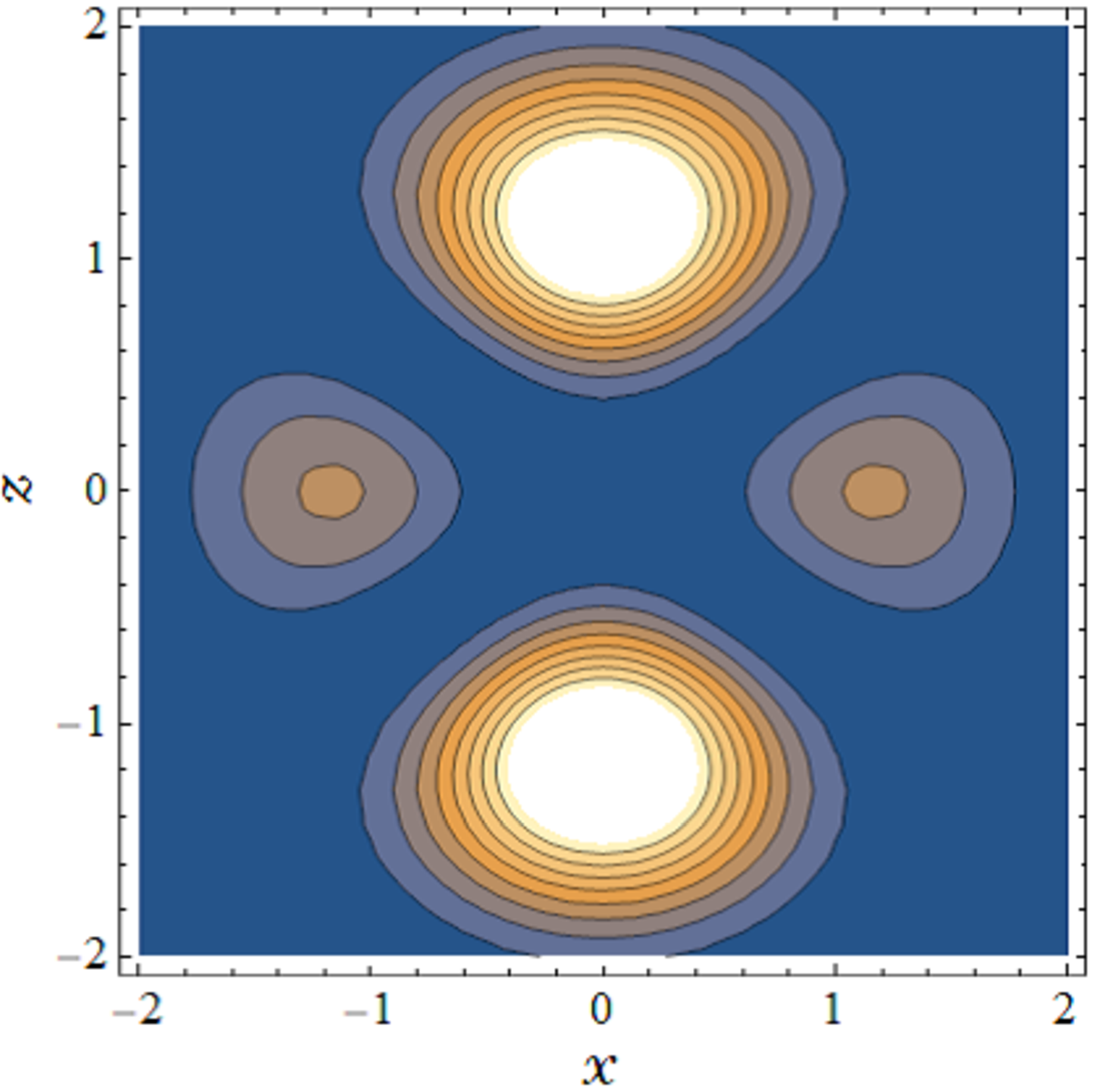}} & 
 \resizebox{70mm}{!}{\includegraphics{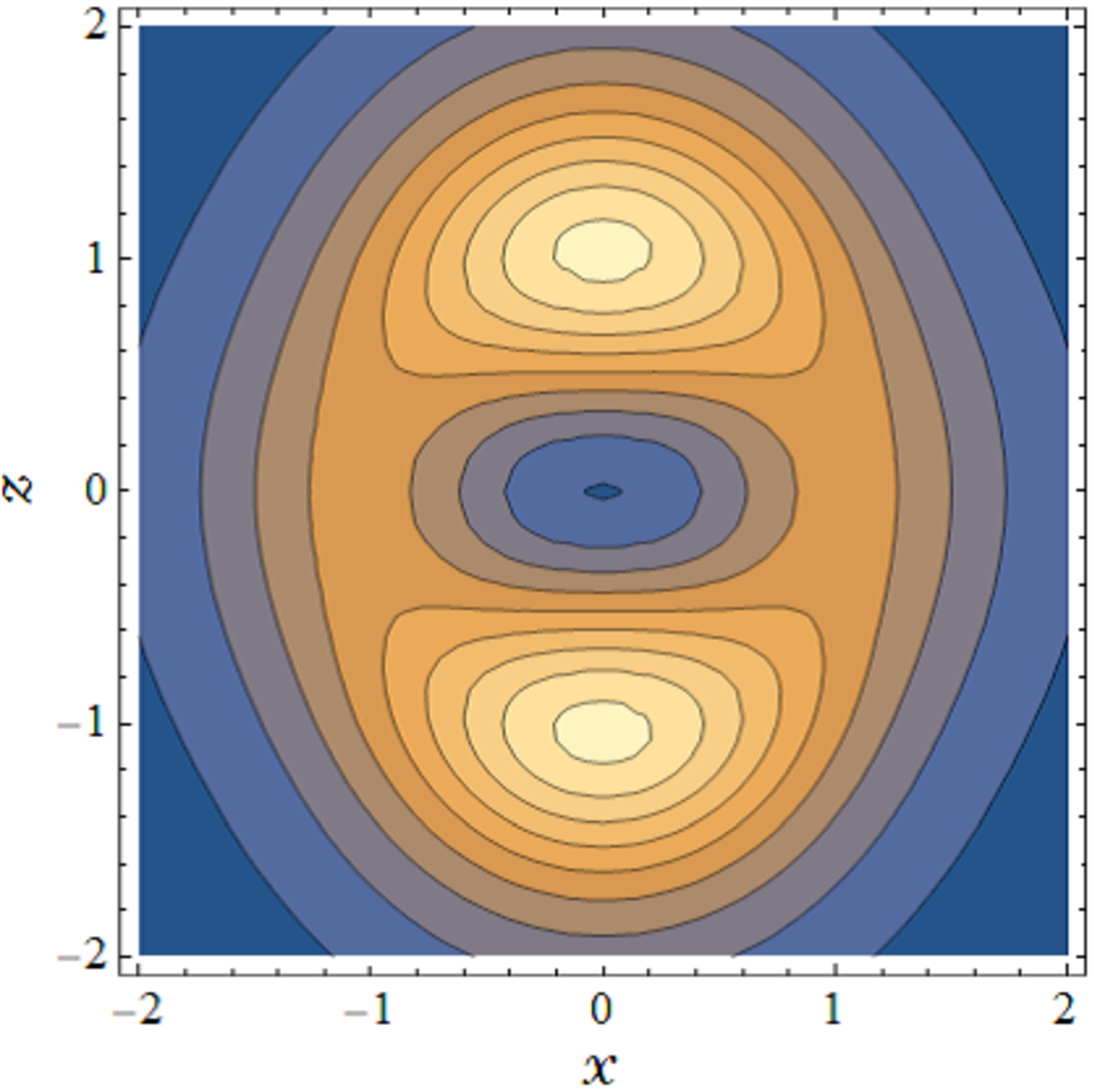}} \\
 \resizebox{70mm}{!}{\includegraphics{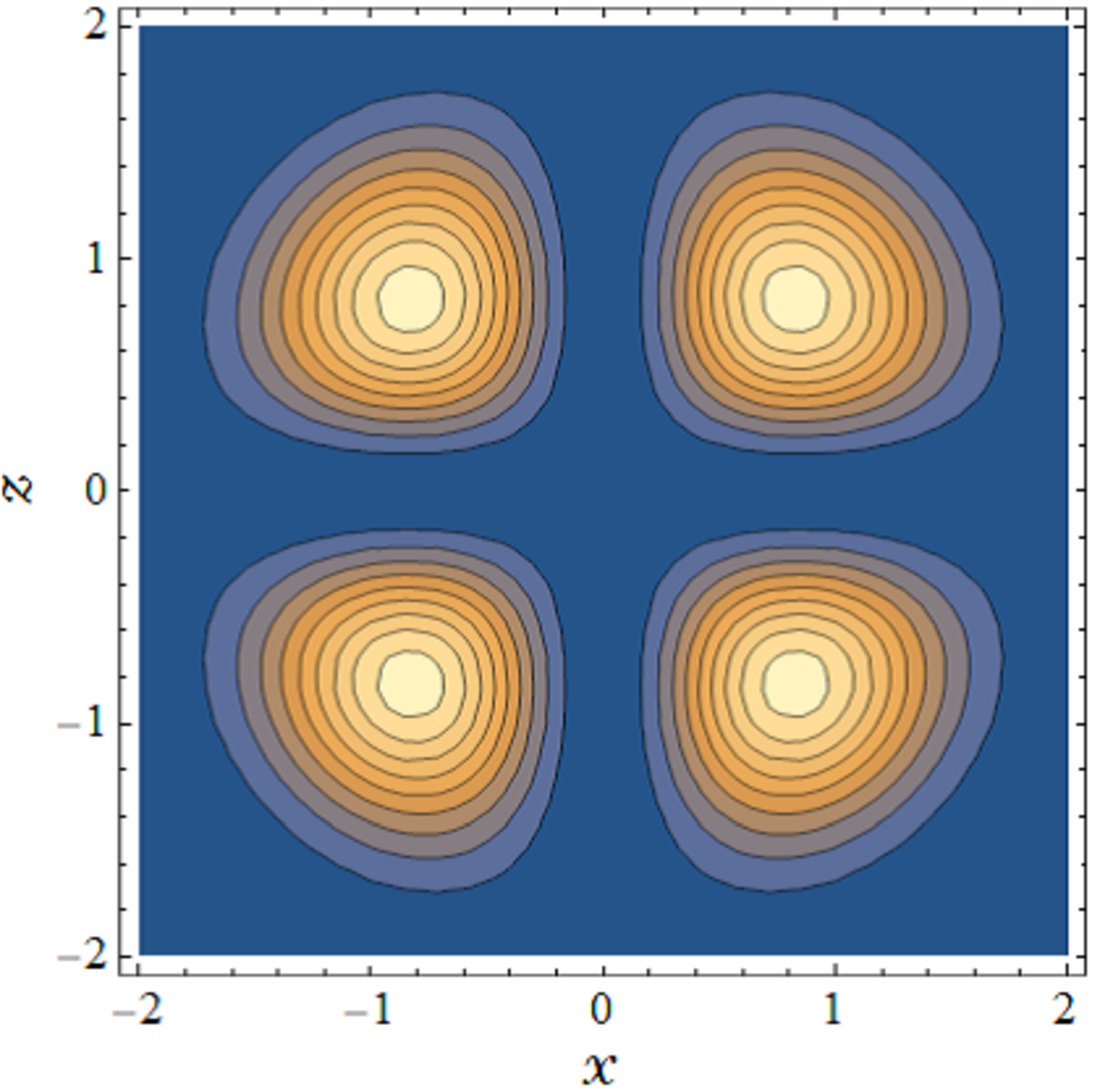}} & 
 \resizebox{70mm}{!}{\includegraphics{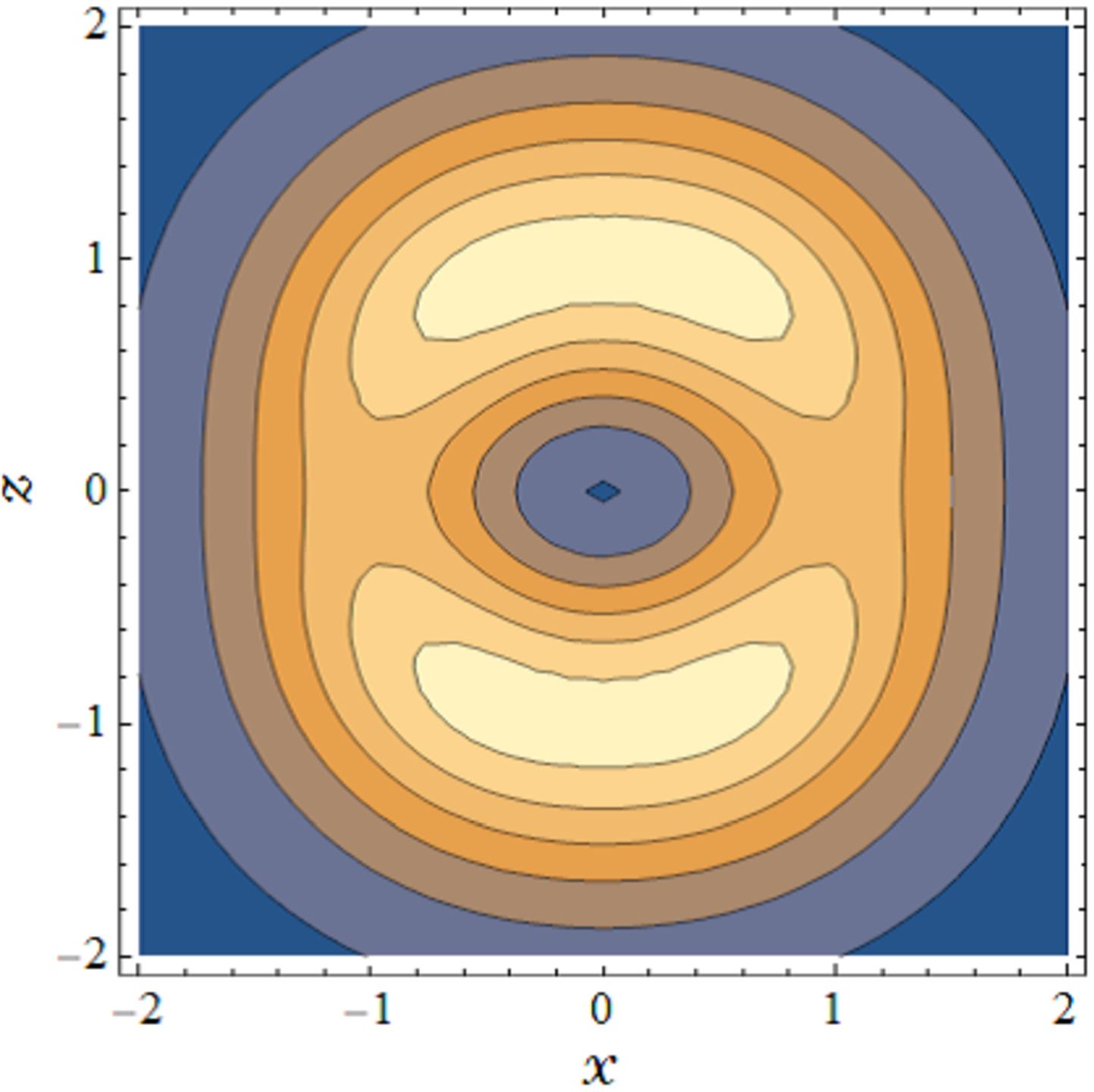}} \\
 \resizebox{70mm}{!}{\includegraphics{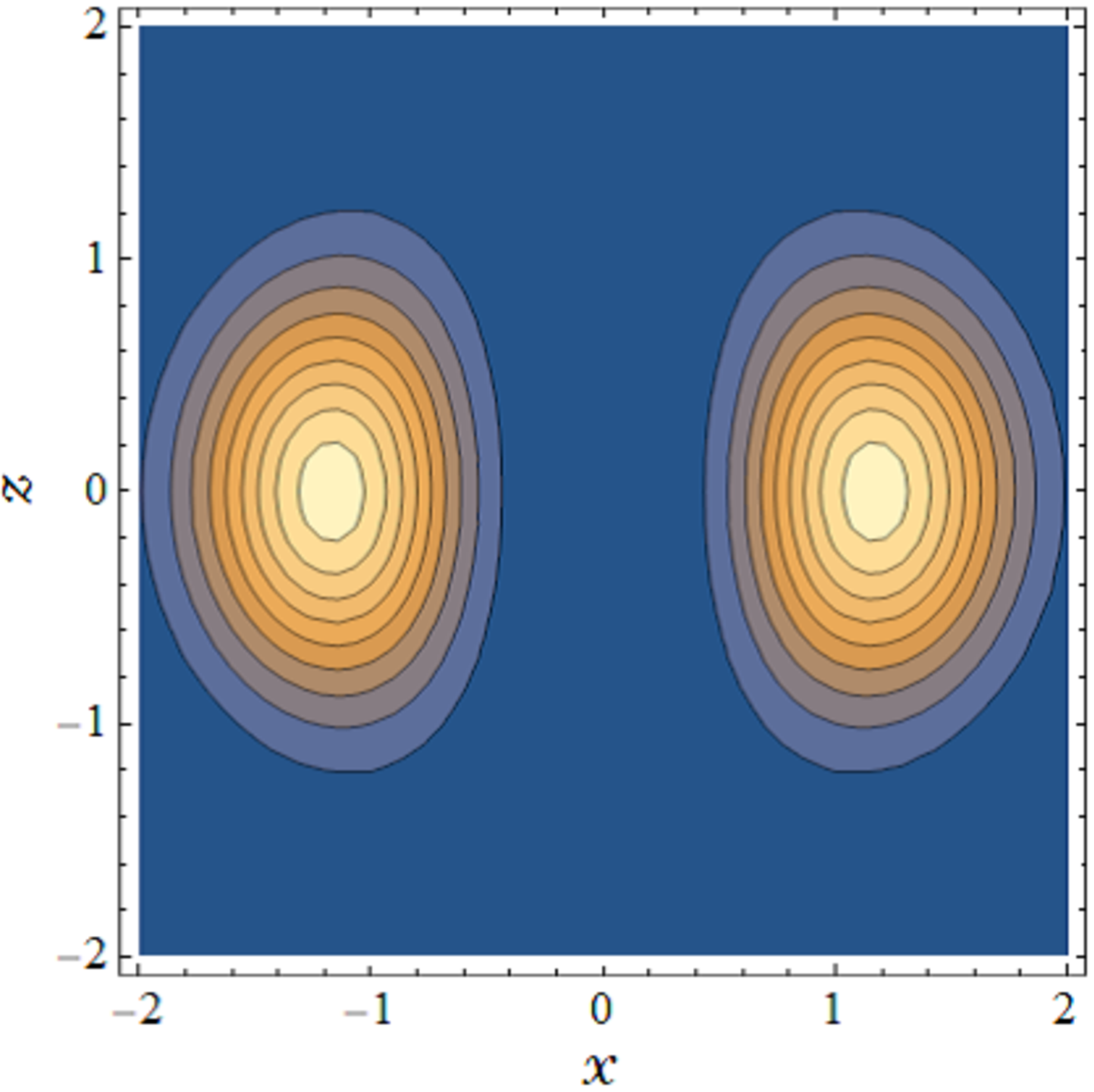}} & 
 \resizebox{70mm}{!}{\includegraphics{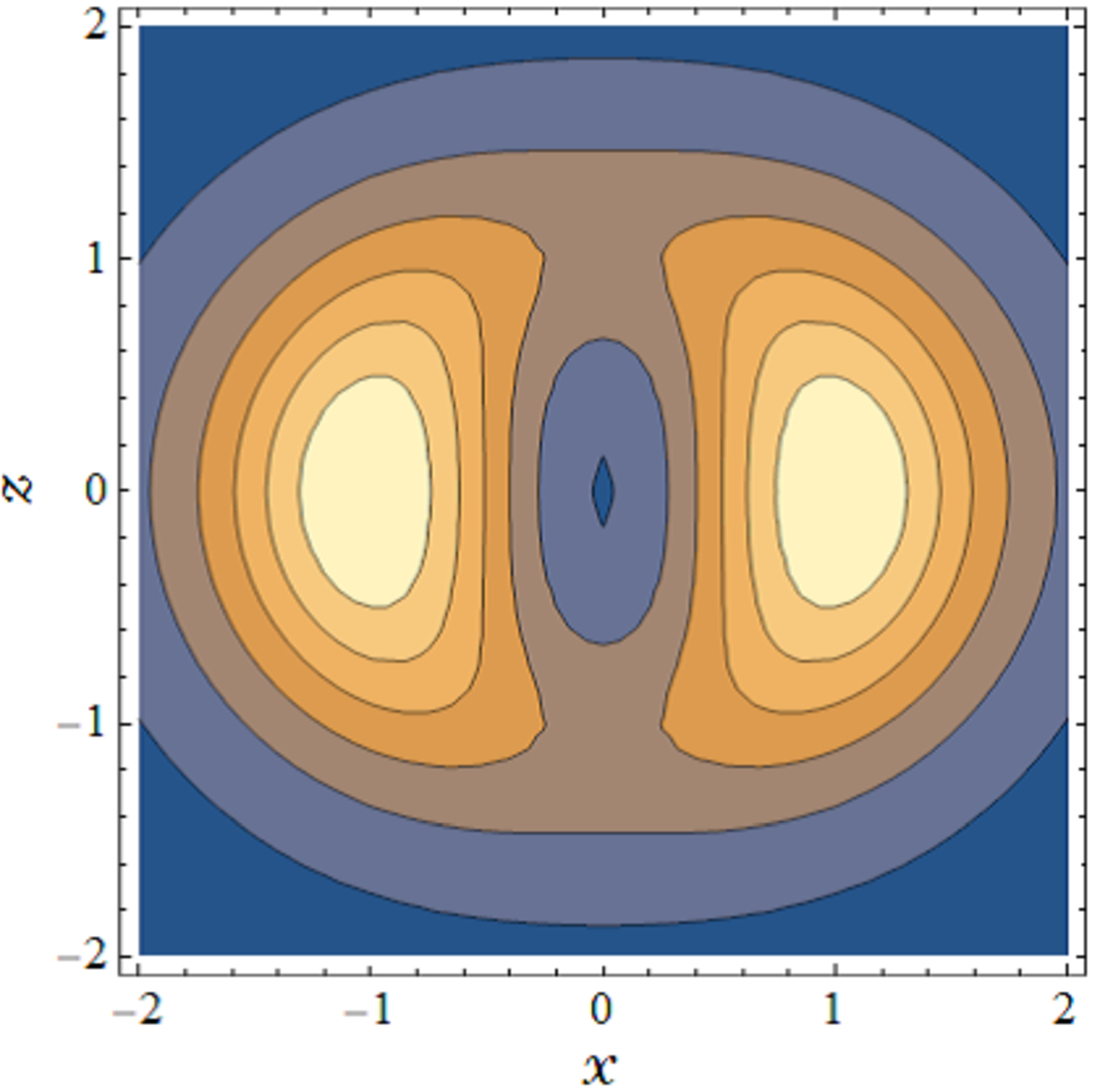}} 
    \end{tabular}
 \caption{The $J=2$ contour plots of the impurity's probability density (left panel) 
 and the real-space excited-boson distributions (right panel):
 the top, middle, and bottom panels are for $|J_z|=0, 1, 2$. 
For other explanations, see the caption in Fig.~\ref{fig4}.}
    \label{fig6}
  \end{center}
\end{figure}

In Fig.~\ref{fig3}, 
we also show the quantum-number distributions of the excited bosons 
$N_{nlm}^{J J_z}$ given by (\ref{Nb1}) for the states of $J=1,2$
as functions of the quantum number $m$ for $l=1,2,3$ and $n=0$,  
with the same parameter set as in Fig.~\ref{fig2}.
%
%
As expected from the angular-momentum conservation and no drag effect, 
i.e., $\langle \hat{M}_z \rangle_{JJ_z}=0$, 
in the present calculation,
all plots in the figures show that 
the distributions for the quantum number $m$ are symmetric about $m=0$.
In order to understand the result,
let's suppose an impurity prepared in the state with a specific value of $J_z(=L_z)$,
which gives a specific direction in the space. 
If the interaction could be turned off between the impurity and surrounding bosons,
the energy of the system should be still degenerate to the value of $J_z$.
However, the presence of the real interaction causes 
the same number of virtual boson excitations 
with the quantum number $-|m|$ and $|m|$
in order to gain the interaction energy by a maximal overlap with the impurity 
(as shown in Fig.~4-6), 
which leads to the vanishing $\langle \hat{M}_z \rangle_{JJ_z}$. 

For a different value of the principal quantum number $n\neq 0$, 
we have confirmed that 
the excited-boson number distributions have the completely same shape as that of $n=0$ 
since distribution shapes are determined by the Clebsch-Gordan coefficients for 
a given set of ($J,J_z,l,m$),
which are independent of $n$,  
but their intensities decrease with increasing $n$. 
The special case is for $J=J_z=0$, 
where the factor $N_{nlm}^{00}$ 
determined from the Clebsch-Gordan coefficients has no $m$ dependence;
their numerical values for $l=1,2,3$ are 
$N_{01m}^{00}=9.76\times10^{-4}$, 
$N_{02m}^{00}=0.38\times10^{-4}$, and
$N_{03m}^{00}=0.024\times10^{-4}$, respectively. 

Finally we show in Figs.~\ref{fig4}-\ref{fig6} the real space distributions (\ref{nbx1}) 
of excited bosons together 
with impurity's probability density obtained from (\ref{ivs1}) 
for the same parameter set as that in Fig.~\ref{fig2}. 
To generate the distributions of excited bosons shown in Fig.~\ref{fig4}-\ref{fig6}, 
we have evaluated (\ref{nbx1}) in the approximation 
of taking quantum numbers up to $(n,l)=(5,5)$ 
in the summation, 
since $f_{nl}$ of higher quantum numbers does not contribute so much 
as the variational parameters (Fig.~\ref{fig2}). 
%
%
%
%
Comparing left and right figures for $J=0,1,2$,  
we can observe that the attractive impurity-boson interaction
has the effect that causes overlaps in their distributions, 
as discussed just above on the quantum-number distributions. 
The impurity's probability is proportional to $|Y_{JJ_z}(\theta,\varphi)|^2$, 
thus the figures clearly exhibit the $s$, $p$, and $d$ orbital shapes for $J=0,1,2$, respectively. 
On the other hand, 
boson's distributions are blurred 
because they always include $l=0$ isotropic contributions 
as shown in (\ref{nbx1}) with variationally-determined weight factor $|f_{nl}|^2$.

\section{Summary and outlook}
In this paper 
we have investigated the ground-state properties of the impurity 
interacting with medium-bosons in spherically symmetric trap potentials, 
when the total angular momentum $(J, J_z)$ are given. 
To this end we have developed a conditional variational method, 
and obtained the ground-state energies, quasi-particle residue of polaron, 
and the quantum-number and real spaces distributions of excited bosons 
for the cases of total angular momenta $J=0,1,2$. 
From theoretical consideration, 
we have found that 
the expectation value
$\langle \hat{J}^2\rangle_{JJ_z}=J(J+1)$ is shared 
by the impurity and the excited bosons as
\beq
\langle \hat{L}^2\rangle_{JJ_z} &=& J(J+1)+\sum'_{n,l} l(l+1)|f_{nl}|^2, \nn
\langle \hat{M}^2\rangle_{JJ_z} &=& \sum'_{n,l} l(l+1)|f_{nl}|^2, 
\nn
\langle 2 \hat{L}\cdot\hat{M}\rangle_{JJ_z} &=& -2\sum'_{n,l} l(l+1)|f_{nl}|^2,
\eeq  
while that of the $z$-th component 
$\langle \hat{J}_z\rangle_{JJ_z}=J_z$ comes from the impurity only: 
\beq
\langle \hat{L}_z \rangle_{JJ_z} = J_z, \quad \langle \hat{M}_z \rangle_{JJ_z}=0, 
\eeq 
which implies no drag effect for the polaron in the spherically-symmetric trap potentials.
We have also made numerical calculations based on the variational method, 
and, 
as shown in Figs.~\ref{fig2}-\ref{fig6}, 
found that the excited bosons are distributed 
so as to make a large overlap with impurity's probability density 
in real and quantum-number spaces 
because of the attractive impurity-boson interaction. 

In the present study 
the excited bosons do not move collectively by themselves \cite{Dalfovo1,Japha1}
since no boson-boson interaction is assumed, 
so they are in purely quantum regime. 
In most of recent experimental researches, 
the Bose polarons are realized in the system of the repulsive boson-boson interactions
where the medium-bosons form a superfluid BEC. 
In order to analyze such cases, 
we need Bogoliubov-de-Gennes-type approaches \cite{Bogol1,Gennes1,Lewenstein4,Nakamura1,Nakamura2,Lampo1} 
beyond the Bogoliubov approximation. 
Such extensions of the present approach for the trapped polaron 
including the boson-boson interactions  
should give more detailed polaron's structures
such as a local depletion of BEC around impurity 
as well as the excitation spectra of the bosonic sector. 

Finally, 
we comment a bit on the possibility of experimental observation of
the finite angular momentum states of the trapped Bose polaron discussed in this paper. 
To our knowledge, 
all experiments have been done with 
axial-symmetric traps for both impurity and medium atoms, 
and no angular momentum is given to the atoms in total. 
To give some finite angular momentum to the system 
in axial symmetric trap potentials, 
we expect that the experimental methods of creating 
a vortex state of the BEC 
can be utilized \cite{Vor1,Vor2}: 
rotating a very dilute impurity-atom gas 
before switching on the interaction with medium bosons, 
and then the whole system, as Bose polaron, 
finally acquires some finite angular momentum.
Furthermore, 
if the axial symmetric trap is deformed adiabatically to the spherical one, 
there remains the state with a finite angular momentum, 
the quantization axis of which should be the same with that of the original axial symmetry. 

For the observation of the angular-momentum distribution of the impurity, 
the photon absorption spectra for excitations for the states with different angular momenta 
can be utilized, 
or the indirect observation of the phase of the impurity's wave function, 
which has been done for the vortex state of the BEC \cite{Vor1}, 
is also an interesting possibility. 
At the moment, 
we have no fixed idea how to give a definite amount of angular momentum, 
but we think that a significant change in boson's distribution 
can be observed with the methods as discussed here.  
Also, in the observation of the excited bosons, 
the photon absorption spectra mentioned above may work out for bosons as well. 
In addition, we think that {\it in-situ} experiments may also work to get images of excited bosons \cite{Yan8,insitu7,insitu8,insitu9,insitu10,insitu11}, 
although it would be a challenge since the total excited-boson number per impurity is quite small. 

Acknowledgments 

We are grateful to Kei Iida, Junichi Takahashi, and Ryosuke Imai for useful discussions on excitation spectra of interacting bosons in inhomogeneous systems, and to Kota Yanase for comments on 
angular momentum structure of excited nuclei. 
E.~N. and  H.~Y. are supported by Grants-in-Aid for Scientific Research 
through Grants No.~17K05445 and No.~18K03501, respectively,  provided by JSPS. 


\begin{thebibliography}{99} 
	%
     %
	\bibitem{Catani1}
	J.~Catani, G.~Lamporesi, D.~Naik, M.~Gring, M.~Inguscio, F.~Minardi, 
	A.~Kantian, and T.~Giamarchi, 
	Phys.~Rev.~A {\bf 85}, 023623 (2012).
	%
	\bibitem{Scelle1}
	R.~Scelle, T.~Rentrop, A.~Trautmann, T.~Schuster, and M.~K.~Oberthaler, 
	Phys.~Rev.~Lett.~{\bf 111}, 070401 (2013). 
	%
	\bibitem{Hohmann1}
	M.~Hohmann, F.~Kindermann, B.~G\"anger, 
	T.~Lausch, D.~Mayer, F.~Schmidt and A.~Widera, 
	EPJ Quantum Technology 2:23,  (2015)
	%
	\bibitem{Compagno1}
	E.~Compagno, G.~De Chiara, D.~G.~Angelakis, and G.~M.~Palma, 
     Scientific Reports vol.~{\bf 7}, 2355 (2017). 
	%
	\bibitem{Jrgensen1}
	N.~B.~J{\o}rgensen, L.~Wacker, K.~T.~Skalmstang, M.~M.~Parish, 
	J.~Levinsen, R.~S.~Christensen, G.~M.~Bruun, J.~J.~Arlt, 
	Phys.~Rev.~Lett.~{\bf 117}, 055302 (2016).
	%
	\bibitem{Hu1}
	M.~-G.~Hu, M.~J.~Van de Graaff, D.~Kedar, J.~P.~Corson, 
	E.~A.~Cornell, and D.~S.~Jin, 
	Phys.~Rev.~Lett.~{\bf 117}, 055301 (2016).
     %
     \bibitem{Rentrop1}
     T.~Rentrop, A.~Trautmann, F.~A.~Olivares, F.~Jendrzejewski, 
     A.~Komnik, and M.~K.~Oberthaler, 
	Phys.~Rev.~X {\bf 6}, 041041 (2016). 
	%
	%
     \bibitem{Froehlich1} 
     B.~Fr{\"o}hlich, M.~Feld, E.~Vogt, M.~Koschorreck, W.~Zwerger, and M.~K{\"o}hl, 
     Phys.~Rev.~Lett.~{\bf 106}, 105301 (2011). 
     %
     \bibitem{Scazza1} 
     F.~ Scazza, G.~Valtolina, P.~Massignan, A.~Recati, A.~Amico, A.~Burchianti, 
     C.~Fort, M.~Inguscio, M.~Zaccanti, and G.~ Roati, 
     Phys.~Rev.~Lett.~{\bf 118}, 083602 (2017). 
     %
     \bibitem{Cetina1} 
     M.~Cetina, M.~Jag, R.~S.~Lous, I.~Fritsche, J.~T.~M.~Walraven, R.~Grimm, 
     J.~Levinsen, M.~M.~Parish, R.~Schmidt, M.~Knap, and E.~Demler, 
     Science {\bf 354}, 96  (2016). 
	%
	%
	\bibitem{PethickSmith1}
	C.~J.~Pethick and H.~Smith, 
	{\it Bose-Einstein Condensation in Dilute Gases} 
	(Cambridge University Press, Cambridge, 2008).
	%
	\bibitem{Pitaevskii1}
	L.~Pitaevskii and S.~Stringari, 
	{\it Bose-Einstein Condensation} (Oxford, New York, 2003).
     %
	%
	%
	\bibitem{Cucchietti1}
	F.~M.~Cucchietti and E.~Timmermans, Phys.~Rev.~Lett.~{\bf 96}, 210401 (2006). 
	%
	\bibitem{Sacha1}
	K.~Sacha and E.~Timmermans, Phys.~Rev.~A {\bf 73}, 063604 (2006).  
	%
	\bibitem{Tempere1}
	J.~Tempere, W.~Casteels, M.~K.~Oberthaler, S.~Knoop, 
	E.~Timmermans, and J.~T.~Devreese, Phys.~Rev.~B {\bf 80}, 184504 (2009). 
	%
	\bibitem{Casteels1}
	W.~Casteels, T.~Van~Cauteren, J.~Tempere, and J.~T.~Devreese, 
	Laser Physics {\bf 21}, 8, pp 1480-1485 (2011).
	%
	\bibitem{Rath1}
	S.~P. Rath and R.~Schmidt, 
	Phys.~Rev.~A, {\bf 88}, 053632 (2013)
	%
	\bibitem{Shashi1}
	A.~Shashi, F.~Grusdt, D.~A.~Abanin, and E.~Demler,
	Phys.~Rev.~A {\bf 89}, 053617 (2014). 
	%
    \bibitem{Li1}
	W.~Li and S.~Das~Sarma,
	Phys.~Rev.~A {\bf 90}, 013618 (2014). 
	%
	\bibitem{Levinsen2}
	J.~Levinsen, M.~M.~Parish, and G.~M.~Bruun, 
	Phys.~Rev.~Lett.~{\bf 115}, 125302 (2015). 
	%
	\bibitem{Dehkharghani5}
	A.~S.~Dehkharghani, A.~G.~Volosniev, and N.~T.~Zinner, 
	Phys.~Rev.~A {\bf 92}, 031601(R) (2015). 
	%
	\bibitem{Ardila2}
	L.~A.~Pe{\~n}a Ardila and S.~Giorgini, 
	Phys.~Rev.~{\bf A} 92, 033612 (2015). 
	%
	\bibitem{Christensen1}
	R.~S.~Christensen, J.~Levinsen, and G.~M.~Bruun, 
	Phys.~Rev.~Lett. {\bf 115}, 160401 (2015)
	%
	\bibitem{Vlietinck1}
	J.~Vlietinck, W.~Casteels, K.~Van~Houcke, J.~Tempere, 
	J.~Ryckebusch, and J.~T.~Devreese,
	New J.~Phys.~{\bf 17},033023 (2015).
	%
	\bibitem{Grusdt1}
	F.~Grusdt, Y.~E.~Shchadilova, A.~N.~Rubtsov, and E.~Demler, 
	Sci.~Rep.~{\bf 5}, 12124 (2015). 
	%
	\bibitem{Grusdt3}
	F.~Grusdt and M.~Fleischhauer, 
	Phys.~Rev.~Lett.~{\bf 116}, 053602 (2016). 
	%
	\bibitem{Shchadilova1}
	Y.~E.~Shchadilova, F.~Grusdt, 
	A.~N.~Rubtsov, and E.~Demler, 
	Phys.~Rev.~A {\bf 93}, 043606 (2016)
     %
     	\bibitem{Shchadilova2}	
	Y.~E.~Shchadilova, R.~Schmidt, F.~Grusdt, 
	and E.~Demler, 
	Phys.~Rev.~ Lett. ~{\bf 117}, 113002 (2016). 
     %
	\bibitem{Grusdt6}
    F.~Grusdt, K.~Seetharam, Y.~Shchadilova, and E.~Demler, 
	Phys.~Rev.~A {\bf 97}, 033612 (2018). 
     %
     \bibitem{Ashida1}
     Y.~Ashida, R.~Schmidt, L.~Tarruell, and E.~Demler
	Phys.~Rev.~B {\bf 97}, 060302(R) (2018). 
     %
     \bibitem{Ardila6}
	L.~A.~Pe{\~n}a Ardila, N.~B. J{\o}rgensen, T. Pohl,  S. Giorgini, G.~M. Bruun, and J.~J. Arlt, 
     arXiv:1812.04609. 
     %
     \bibitem{Nielsen3}
     K.~K.~Nielsen,	L.~A.~Pe{\~n}a Ardila, G.~M. Bruun, and T.~Pohl, 
     arXiv:1806.09933. 
     %
     \bibitem{Mistakidis8}
     Effective One-Body Approach to Impurities in One-Dimensional Trapped Bose Gases, 
     S.~I.~Mistakidis, A.~G.~Volosniev, N.~T.~Zinner, P.~Schmelcher , 
     arXiv:1809.01889
     %
     \bibitem{Mistakidis9}
     Quench Dynamics and Orthogonality Catastrophe of Bose Polarons, 
      S.~I.~Mistakidis, G.~C.~Katsimiga, G.~M.~Koutentakis, Th.~Busch, P.~Schmelcher, 
     arXiv:1811.10702
	%
	\bibitem{polaronreview1}
	See, for instance, 
	A.~S.~Alexandrov, J.~T.~Devreese, 
	{\it Advances in Polaron Physics},
	Springer Series in Solid-State Sciences Vol. 159, (Springer, 2009). 
	\bibitem{Landau1}
	L.~D.~Landau, Phys.~Z.~Sowjetunion {\bf 3}, 664 (1933);
	L.~Landau and S.~Pekar, J.~Exptl.~Theor.~Phys. {\bf 18}, 419 (1948);
	S.~Pekar, J.~Exptl.~Theor.~Phys. {\bf 19}, 796 (1949). 
	%
	\bibitem{FPZ1}
	H.~Fr\"ohlich, {\it Theory of Dielectrics}, 
	(Clarendon Press, Oxford, 1949);
	H.~Fr\"ohlich, H.~Pelzer, and S.~Zienau, Phil.~Mag. {\bf 41}, 221 (1950);
	H.~Fr\"ohlich, Adv.~Phys. {\bf 3}, 325 (1954). 
	%
	%
	\bibitem{Chevy1}
	F.~Chevy, 
	Phys.~Rev.~A {\bf 74}, 063628 (2006); 
	%
      Unitary polarized Fermi gases, p.~607 in {\it Ultra-Cold Fermi Gases}, 
      Eds. M.~Inguscio, W.~Ketterle, C.~Salomon,   (IOS Press, Amsterdam, 2007). 
	%
     \bibitem{Massignan1}
     P.~Massignan, G.~M.~Bruun, and H.~T.~C.~Stoof
     Phys.~Rev.~A {\bf 78}, 031602(R) (2008). 
     %
	\bibitem{Schirotzek1}
	A.~Schirotzek, C.~-H.~Wu, A.~Sommer, and M.~~W.~Zwierlein, 
	Phys.~Rev.~Lett.~{\bf 102}, 230402 (2009). 
      %
     \bibitem{Ku1}
     M.~Ku, J.~Braun, and A.~Schwenk, 
	Phys.~Rev.~Lett.~{\bf 102}, 255301 (2009). 
	%
	\bibitem{Schmidt1}
	R.~Schmidt and T.~Enss, 
	Phys.~Rev.~A {\bf 83}, 063620 (2011).  
	%
	\bibitem{Kohstall1}
	C.~Kohstall, M.~Zaccanti, M.~Jag, A.~Trenkwalder, P.~Massignan, 
	G.~M.~Bruun, F.~Schreck, and R.~Grimm, 
	Nature~{\bf 485}, 615-618 (2012). 
	%
	\bibitem{Koschorreck1}
	M.~Koschorreck, D.~Pertot, E.~Vogt, B.~Fr\"ohlich, 
	M.~Feld, and M.~K\"ohl, 
	Nature {\bf 485}, 619 (2012).  
     %
     \bibitem{Schmidt6}
     R.~Schmidt, T.~Enss, V.~Pietil\"a, and E.~Demler, 
	Phys.~Rev.~A {\bf 85}, 021602(R)  (2012). 
	%
	\bibitem{Vlietinck2}
	J.~Vlietinck, J.~Ryckebusch, and K.~Van~Houcke, 
	Phys.~Rev.~B {\bf 87}, 115133 (2013). 
	%
     \bibitem{Trefzger1}
     C.~Trefzger and Y.~ Castin, 
     Europhysics Letters {\bf 104}, 50005 (2013). 
     %
     \bibitem{Trefzger2}
     C.~Trefzger and Y.~ Castin, 
	Phys.~Rev.~A {\bf 90}, 033619 (2014). 
     %
	\bibitem{Massignan5}
	P.~Massignan, M.~Zaccanti, and G.~M.~Bruun, 
	Reports on Progress in Physics, {\bf 77}, 034401, (2014). 
     %
     \bibitem{Lan1}
     Z.~Lan and C.~ Lobo, 
     J.~Indian Inst.~Sci.|{\bf 94}, 179 (2014)
     %
     \bibitem{Lan2}
     Z.~Lan and C.~ Lobo, 
	Phys.~Rev.~A {\bf 92}, 053605 (2015)
	%
     \bibitem{Nur1}
     F.~N.~\"Unal, B.~Het\'enyi, and M.~\"O.~Oktel, 
	Phys.~Rev.~A {\bf 91}, 053625 (2015). 
	%
	\bibitem{Yi1}
	W.~Yi and X.~Cui
	Phys.~Rev.~A {\bf 92}, 013620 (2015).
     %
	\bibitem{Parish5} 
	M.~M.~Parish and J.~Levinsen, 
	Phys.~Rev.~B {\bf 94}, 184303 (2016). 
	%
     \bibitem{Levinsen4}
     J.~Levinsen, P.~Massignan, S.~Endo, M.~M.~Parish, 
     J.~Phys.~B: At.~Mol.~Opt.~Phys.~{\bf 50}, 072001 (2017). 
     %
     \bibitem{Kamikado1}
     K.~Kamikado, T.~ Kanazawa, and S.~ Uchino, 
	Phys.~Rev.~A {\bf 95}, 013612 (2017). 
     %
     \bibitem{Kain1}
     B.~Kain and H.~Y.~Ling, 
	Phys.~Rev.~A {\bf 96}, 033627 (2017)
     %
     \bibitem{Schmidt5} 
     R.~Schmidt, M.~Knap, D.~A.~Ivanov, J.-S.~ You, M.~ Cetina, E.~ Demler, 
     	Rep.~Prog.~Phys.~{\bf 81}, 024401 (2018). 
     %
     \bibitem{Mistakidis10} 
     Repulsive Fermi Polarons and Their Induced Interactions in Binary Mixtures of    Ultracold Atoms, 
     S.~I.~Mistakidis, G.~C.~Katsimiga, G.~M.~Koutentakis, P.~Schmelcher, 
 	arXiv:1808.00040. 
     %
	%
     \bibitem{Levinsen5} 
    J.~Levinsen, P.~Massignan, F.~Chevy, and C.~Lobo, 
    Phys.~ Rev.~ Lett.~{\bf 109}, 075302 (2012). 
     %
     %
     \bibitem{Deng1} 
     T.~-S.~Deng, Z.~-C.~Lu, Y.~-R.~Shi, J.~-G.~Chen, W.~Zhang, and W.~Yi, 
	Phys.~Rev.~A {\bf 97}, 013635 (2018). 
    %
    \bibitem{Kain3}
    B.~Kain and H.~Y.~Ling, 
	Phys.~Rev.~A {\bf 89}, 023612 (2014).  
    %
    \bibitem{Ardila4}
     Ground-state properties of Dipolar Bose polarons, 
     L.~A.~Pe{\~n}a Ardila  and T.~Pohl, 
     arXiv:1804.06390
     %
     %
     \bibitem{Tajima1}
     H.~Tajima and S.~Uchino, 
     New~J.~Phys. {\bf 20}, 073048 (2018).
     %
     \bibitem{Yan8}
     Boiling a Unitary Fermi Liquid, 
     Z.~Yan, P.~B.~Patel, B.~Mukherjee, R.~J.~Fletcher, J.~Struck, and M.~W.~Zwierlein, 
     arXiv:1811.00481
     %
     \bibitem{Tajima2}
     Thermal crossover, transition, and coexistence in Fermi polaronic spectroscopies, 
     H.~Tajima and S.~Uchino, 
     arXiv:1812.05889
     %
     \bibitem{Levinsen8}
     J.~Levinsen, M.~M.~Parish, R.~S.~Christensen, J.~J.~Arlt, and G.~M.~Bruun, 
     Phys.~Rev.|A {\bf 96}, 063622 (2017).
     %
     \bibitem{Guenther8}
     N.-E.~Guenther,  P.~Massignan, M.~Lewenstein, and G.~M. Bruun, 
     Phys.~Rev.~Lett.{\bf 120}, 050405 (2018). 
     %
	%
	\bibitem{Bruderer1}
	M.~Bruderer, Alexander~Klein, Stephen~R.~Clark, and Dieter~Jaksch, 
	Phys.~Rev.~A {\bf 76}(R):011605 (2007); 
	New~J.~Phys.~{\bf 10}, 033015 (2008). 
     %
     \bibitem{Nakano2}
      E.~Nakano and H.~Yabu, 
	Phys.~Rev.~B {\bf 93}, 205144 (2016). 
     %
     %
	%
     \bibitem{Rowe1} 
      For instance, see, 
      David J.~Rowe, {\it Nuclear collective motion : models and theory} 
      (World Scientific, New Jersey, 2010)
     %
     \bibitem{Inglis1} 
	D.~R.~Inglis, Phys.~Rev.~{\bf 96}, 1059 (1954) ; ~{\bf 96}, 701 (1955).
	%
	\bibitem{Thouless1}
	D.~J.~Thouless and J.~G.~Valatin, 
	Nucl.~Phys.~{\bf 31}, 211-230 (1962). 
	%
     %
	\bibitem{Lemeshko3}
	R.~Schmidt and M.~Lemeshko, 
     Phys.~Rev.~Lett.~{\bf 114}, 203001 (2015). 
     %
	\bibitem{Lemeshko4}
	R.~Schmidt and M.~Lemeshko, 
	Phys.~Rev.~X {\bf 6}, 011012 (2016). 
	%
	\bibitem{Lemeshko7}
     E.~Yakaboylu, B.~Midya, A.~Deuchert, N.~Leopold, and M.~Lemeshko
     Phys.~Rev.~B {\bf 98}, 224506 (2018). 
     %
     %
	%
     \bibitem{NYI1} 
     E.~Nakano, H.~Yabu, and K.~Iida, 
     Phys.~Rev.~A {\bf 95}, 023626 (2017)
	%
     %
     \bibitem{Rose1} 
     M.~E.~Rose, 
     {\it Elementary Theory of Angular Momentum} (Dover Publications, 2011). 
     %
     \bibitem{Edmonds1}
     A.~R.~Edmonds, 
     {\it Angular Momentum in quantum mechanics} (Princeton University Press, New Jersey, 1957).
     %
	%
	\bibitem{LLP1}
	T.~D.~Lee, F.~E.~Low, and D.~Pines, 
	Phys.~Rev.~{\bf 90}, No.2, 297-302 (1953). 
     %
	\bibitem{Schweber1}
	S.~S.~Schweber, {\it An Introduction to Relativistic Quantum Field Theory}, 
	(Dover Publications, New York, 2005). 
     %
	%
	%
	%
	%
	%
	%
	\bibitem{Dalfovo1}
	F.~Dalfovo, S.~Giorgini, L.~P.~Pitaevskii, and S.~Stringari, 
	Rev.~Mod.~Phys.~{\bf 71}, 463 (1999). 
	%
	\bibitem{Japha1}
	Y.~Japha and Y.~B.~Band, 
	Phys.~ Rev.~A {\bf 84}, 033630 (2011). 
     %
	%
	\bibitem{Bogol1}
	N.~N.~Bogoliubov, 
	J.~Phys.~(Moscow) {\bf 11}, 32 (1947).
	%
	\bibitem{Gennes1}
	P.~G.~de~Gennes, 
	{\it Superconductivity of Metals and Alloys} (Benjamin, New York, 1966). 
	%
	\bibitem{Lewenstein4}
	M.~Lewenstein and L.~You, 
	Phys.~Rev.~Lett. {\it 77}, 3489 (1996).
	%
	\bibitem{Nakamura1}
	Y.~Nakamura, J.~Takahashi, and Y.~Yamanaka, 
	Phys.~ Rev.~A {\bf 89} 013613 (2014). 
	%
	\bibitem{Nakamura2}
	Y.~Nakamura, T.~Kawaguchi, Y.~Torii, and Y.~Yamanaka, 
	Annals of Physics {\bf 376}, 484 (2017).  
	%
    \bibitem{Lampo1}	
	A.~Lampo, C.~Charalambous, M.~A.~Garcia-March, and M.~ Lewenstein, 
	arXiv:1803.08946. 
     %
	%
    \bibitem{Vor1}
	M.~R. Matthews, B.~P. Anderson, P.~C. Haljan, D.~S. Hall, C.~E. Wieman, and E.~A. Cornell,
    Phys. Rev. Lett. {\bf 83}, 2498 (1999).
	%
	\bibitem{Vor2}
	K.~W. Madison, F. Chevy, W. Wohlleben, and J. Dalibard, 
	Phys. Rev. Lett. {\bf 84}, 806 (2000);
	J.~R. Abo-Shaeer, C. Raman, J.~M. Vogels, W. Ketterle, Science {\bf 292}, 476 (2001).
     %
	%
    %
	%
	\bibitem{insitu7}
     Y.~-I.~Shin, C.~H.~Schunck, A.~Schirotzek, and W.~Ketterle,
      Nature {\bf 451}, 689-693 (2008).
     %
	\bibitem{insitu8}
     W.~S.~Bakr, J.~I.~Gillen, A.~Peng, S.~F{\"o}lling, and M.~Greiner, 
     Nature {\bf 462}, 74-77 (2009). 
     %
	\bibitem{insitu9}
     J.~F.~Sherson, C.~Weitenberg, M.~Endres, M.~Cheneau, I.~Bloch, and S.~Kuhr, 
      Nature {\bf 467}, 68-72 (2010). 
     	%
	\bibitem{insitu10}
     M.~Horikoshi, S.~Nakajima, M.~Ueda, and T.~Mukaiyama,
     Science {\bf 327}, 442 (2010).
     	%
	\bibitem{insitu11}
     M.~J.~H.~Ku, A.~T.~Sommer, L.~W.~Cheuk, and M.~W.~Zwierlein, 
     Science {\bf 335}, 563 (2012).
     	%
\end{thebibliography}

\appendix
\section{Expectation values of operators by the coherent states}
In this appendix we present the expectation values of the gauge-transformed operators 
$S^{-1}\hat{J}^2S$ 
and $S^{-1}{\mathcal H} S$, 
which are defined in (\ref{traJsq}) and (\ref{traH}),  
with respect to the coherent state (\ref{coh1}). 
Using the expectation values of $\hat{M}_i$ and $\hat{L}_i$:
\beq
\left\langle \hat{M}_{i}\right\rangle_b &=& 0,  \\
\left\langle \hat{M}_{\pm1}\hat{M}_{\mp1}\right\rangle_b 
&=& 
\Braket{l,0|\hat{\mathcal{L}}_{\pm1}\hat{\mathcal{L}}_{\mp1} |l,0} |f_{nl}|^2
=
-\frac{l(l+1)}{2} |f_{nl}|^2, 
\eeq
and 
$\left\langle \hat{M}_{i}\hat{M}_{j}\right\rangle_b=0$ 
for the other combinations of $i$ and $j$, 
we obtain the expectation value of the shift operator: 
\beq
\langle \hat{O}_L \rangle_b 
&=&
\langle S^{-1}(\hat{L}^2S)\rangle_b
\nn
&=&
\left\langle 
-\frac{\cot\theta}{\sqrt{2}}\lk \hat{M}_{-1}
+\hat{M}_{+1}\rk-\frac{1}{2}\lk \hat{M}_{-1}+\hat{M}_{+1}\rk^2
\right.
\nn
&&
\qquad +
\left.
\frac{1}{\sin^2\theta} 
\ldk \cos\theta \hat{M}_0
-\sin\theta \frac{1}{\sqrt{2}}\lk \hat{M}_{-1}-\hat{M}_{+1}\rk \rdk^2 
\right\rangle_b
\nn
&=&
-\frac{1}{2}
\left\langle \lk \hat{M}_{-1}+\hat{M}_{+1}\rk^2\right\rangle_b 
+\frac{1}{2}
\left\langle 
\lk \hat{M}_{-1}-\hat{M}_{+1}\rk^2 
\right\rangle_b
\nn
&=&
l(l+1) |f_{nl}|^2.
\eeq
Then the expectation value of the transformed squared total angular momentum operators
becomes  
\beq
\left\langle S^{-1}\hat{J}^2S \right\rangle_b
&=& 
\left\langle \hat{L}^2 \right\rangle_b
+
\left\langle \hat{M}^2\right\rangle_b
+
\left\langle \hat{O}_L \right\rangle_b
+2 \sum_{i,j} D^1_{i,j}(\varphi, \theta, 0)
\left\langle \hat{M}_j ^\dagger
\ldk 
S^{-1}(\hat{L}_i S)+\hat{L}_i
\rdk
\right\rangle_b
\nn
&=& 
\hat{L}^2
+
l(l+1)|f_{nl}|^2  
+
l(l+1)|f_{nl}|^2
+2 \sum_{i,j} D^1_{i,j}(\varphi, \theta, 0)
\left\langle \hat{M}_j ^\dagger
S^{-1}(\hat{L}_i S)
\right\rangle_b
\nn
&=& 
\hat{L}^2
+
2l(l+1)|f_{nl}|^2 
-2 \sum_{i=0, \pm1} 
\Big\{
D^1_{i,-1}(\varphi, \theta, 0)
\left\langle \hat{M}_{+1}
S^{-1}(\hat{L}_i S)
\right\rangle_b
\nn
&&\qquad\qquad\qquad\qquad\qquad\qquad\qquad
+
D^1_{i,+1}(\varphi, \theta, 0)
\left\langle \hat{M}_{-1}
S^{-1}(\hat{L}_i S)
\right\rangle_b
\Big\}
\nn
&=& 
\hat{L}^2, 
\eeq
Finally, 
we obtain the expectation value of the transformed Hamiltonian:
\begin{eqnarray}
\langle S^{-1} {\mathcal H}S \rangle_b
&=&
H_{ho}(\br) +\frac{l(l+1)|f_{nl}|^2}{2m_fr^2}+ E^b_{00} N_0 
+ \sum_{n,l}'E_{nl}^b 
|f_{nl}|^2
+g N_0 |\phi_0^b(\br)|^2
\nonumber\\
&&+g\frac{\sqrt{N_0}}{4\pi} \sum_{n,l}'
\sqrt{2l+1} R_{00}^{b}(r) R_{nl}^b(r) \ldk f_{nl} +f^*_{nl} \rdk. 
\end{eqnarray}

\section{The variational energy functional in terms of dimensionless variables}
Here we present the coefficients appearing 
in the functionals (\ref{functio1}) and (\ref{functio2}).
The functional $G[F_{0J};F_{1J}]$ is expanded as,
\beq
G[F_{0J};F_{1J}] &:=& \int_\br \frac{1}{r^2}
\left|\sum_{n=0,1} F_{nJ} \phi^I_{nJJ_z}(\br)\right|^2  
\nn 
&=&
|F_{0J}|^2 G^0_J
+ F_{0J}^*F_{1J} G^c_J
+ F_{1J}^*F_{0J} {G^c}^*_J
+|F_{1J}|^2 G^1_J,  
\eeq
where the factors $G^0_J$, $G^1_J$, $G^c_J$ 
is given as 
$G^0_J = G^1_J=G^c_J \sqrt{J+\frac{3}{2}}
=\frac{\omega_I m_I}{J+\frac{1}{2}}$.

The another functional $H[F_{0J};F_{1J}]_{nl}$ is represented as
%
\beq
H[F_{0J};F_{1J}]_{nl} &:=& \int_\br R_{00}^{b}(r) R_{nl}^b(r)
\left|\sum_{n=0,1} F_{nJ} \phi^I_{nJJ_z}(\br)\right|^2
\nn 
&=&
|F_{0J}|^2 H_{nlJ}^0
+ F_{0J}^*F_{1J} H_{nlJ}^c
+ F_{1J}^*F_{0J} {H_{nlJ}^c}^*
+|F_{1J}|^2H_{nlJ}^1, 
\eeq 
%
where  
\beq
H_{nlJ}^0
&=& 
\int_0^\infty {\rm d}r r^2 
R_{00}^{b}(r) R_{nl}^b(r) R^I_{0J}(r)^2
\nn
&=&  
\lk m_I \omega_I\rk^{\frac{3}{2}}
{\mathcal N}_{00}{\mathcal N}_{nl}{\mathcal N}_{0J}^2
Q^{2J} 
\int_0^\infty {\rm d}x x^{2+l+2J} 
\, e^{-\lk 1+Q^2 \rk x^2} L_n^{\lk\frac{1+2l}{2}\rk}\lk x^2\rk
\nn 
&=& 
\lk m_I \omega_I\rk^{\frac{3}{2}}
{\mathcal N}_{00}{\mathcal N}_{nl}{\mathcal N}_{0J}^2
Q^{2J}(1+Q^2)^{-\frac{3}{2}-J-\frac{l}{2}}
\frac{\Gamma[\frac{3+2J+l}{2}]\Gamma[\frac{3+2l+2n}{2}]}
{2\Gamma[\frac{3+2l}{2}]n!} \nn
&&\qquad \times F\left(\frac{3+2J+l}{2},-n,\frac{3+2l}{2},\frac{1}{1+Q^2}\right), 
\\ 
H_{nlJ}^c
&=&  
\int_0^\infty {\rm d}r r^2 
R_{00}^{b}(r) R_{nl}^b(r) R^{I}_{0J}(r) R^I_{1J}(r)
\nn
&=&  
\lk m_I \omega_I\rk^{\frac{3}{2}}
{\mathcal N}_{00}{\mathcal N}_{nl}{\mathcal N}_{0J}{\mathcal N}_{1J} 
Q^{2J} \nn
&&\qquad \times \int_0^\infty {\rm d}x x^{2+l+2J} 
\, e^{-\lk 1+Q^2 \rk x^2}  L_n^{\lk\frac{1+2l}{2}\rk}\lk x^2\rk 
L_1^{\lk\frac{1+2J}{2}\rk}\lk Q^2 x^2\rk, 
\nn 
\\ 
H_{nlJ}^1
&=& 
\int_0^\infty {\rm d}r r^2 
R_{00}^{b}(r) R_{nl}^b(r) R^I_{1J}(r)^2
\nn
&=&  
\lk m_I \omega_I\rk^{\frac{3}{2}}
{\mathcal N}_{00}{\mathcal N}_{nl}{\mathcal N}_{1J}^2
Q^{2J} \nn
&&\qquad \times \int_0^\infty {\rm d}x x^{2+l+2J} 
\, e^{-\lk 1+Q^2 \rk x^2}
L_n^{\lk\frac{1+2l}{2}\rk}\lk x^2\rk\, 
\left\{ 
L_1^{\lk\frac{1+2J}{2}\rk}\lk Q^2 x^2\rk 
\right\}^2, 
\eeq
where $Q=\sqrt{\frac{m_I \omega_I}{m_b \omega_b} }$.
The $\Gamma(z)$ and $F(a,b,c,z)$ in the above formulas represent 
the gamma and Gauss's hypergeometric functions, respectively. 
We have shown the analytic expression only for $H_{nlJ}^0$, 
but the remaining factors $H_{nlJ}^c$ and $H_{nlJ}^1$ also 
have similar analytic expressions,
which are not presented here because they are lengthy and cumbersome. 

In terms of dimensionless variables, 
the polaron binding energy (the energy shift (\ref{shift1}) ), 
defined by the energy difference of the systems  
with and without the impurity-medium interaction is given by
\beq
\frac{\Delta E[F_{1J}]}{\omega_b} 
&=& 
2\beta F_{1J}^2
+\frac{N_0}{2} \lk 1+\alpha \rk 
\alpha^{\frac{1}{2}}
\beta^{\frac{3}{2}}
\gamma 
\ldk \tilde{H}_{00J}^0+ 2F_{1J} \sqrt{1-F_{1J}^2}\tilde{H}_{00J}^c+F_{1J}^2
\lk \tilde{H}_{00J}^1- \tilde{H}_{00J}^0\rk \rdk
\nn 
&&
-\frac{N_0}{4}\lk 1+\alpha \rk^2 
\alpha
\beta^3
\gamma^2 \nn
&&\qquad \times \sum_{n,l}'
(2l+1)  
\frac{\ldk \tilde{H}_{nlJ}^0+ 2F_{1J} \sqrt{1-F_{1J}^2}\tilde{H}_{nlJ}^c+F_{1J}^2
\lk \tilde{H}_{nlJ}^1- \tilde{H}_{nlJ}^0\rk \rdk^2}
{\frac{3+2l+4n}{2}+\frac{l(l+1)}{2\lk J+\frac{1}{2}\rk}\beta 
\ldk 1+ 2F_{1J} \sqrt{1-F_{1J}^2}\lk J+\frac{3}{2}\rk^{-1/2} \rdk}, 
\nn 
&=& 
2\beta F_{1J}^2
-\frac{3\sqrt{N_0}}{2} \bar{f}_{00}[F_{1J}]
\nn 
&&
-
\sum_{n,l}'
\bar{f}_{nl}[F_{1J}]^2
\ldk \frac{3+2l+4n}{2}+\frac{l(l+1)}{2J+1}\beta 
	\lk 1+ 2\frac{F_{1J} \sqrt{1-F_{1J}^2}}{\sqrt{J+\frac{3}{2}}}\rk\rdk, 
\nn 
\eeq 
where we have used the formal solution (\ref{fnl1}) :
\beq
 \bar{f}_{nl}[F_{1J}]&=&
-\frac{\sqrt{N_0}}{2}\lk 1+\alpha \rk \alpha^{\frac{1}{2}}\beta^{\frac{3}{2}}\gamma \nn
&& \times \sqrt{2l+1}   
\frac{ \tilde{H}_{nlJ}^0+ 2F_{1J} \sqrt{1-F_{1J}^2}\tilde{H}_{nlJ}^c+F_{1J}^2
	\lk \tilde{H}_{nlJ}^1- \tilde{H}_{nlJ}^0\rk}
{\frac{3+2l+4n}{2}+\frac{l(l+1)}{2\lk J+\frac{1}{2}\rk}\beta 
	\ldk 1+ 2F_{1J} \sqrt{1-F_{1J}^2}\lk J+\frac{3}{2}\rk^{-1/2} \rdk}, 
\eeq
%
with $\tilde{H}_{nlJ}^0\equiv H_{nlJ}^0/(m_I\omega_I)^{3/2}$ and so on, 
$\alpha \equiv m_I/m_b$, 
$\beta\equiv \omega_I/\omega_b$, and $\gamma\equiv a (m_b\omega_b)^{1/2}$ via 
\beq
\frac{g}{\omega_b} \lk m_I \omega_I\rk^{\frac{3}{2}} 
&=& 
\frac{2\pi a\lk m_b+m_I\rk }{m_b m_I} \frac{\lk m_I \omega_I\rk^{\frac{3}{2}}}{\omega_b} 
=
2\pi \lk 1+\frac{m_I}{m_b } \rk \frac{\omega_I}{\omega_b} 
a \lk m_I \omega_I\rk^{\frac{1}{2}}
\nn
&=&
2\pi \lk 1+\frac{m_I}{m_b } \rk
\lk \frac{m_I}{m_b} \rk^{\frac{1}{2}}
\lk \frac{\omega_I}{\omega_b} \rk^{\frac{3}{2}}
a \lk m_b \omega_b\rk^{\frac{1}{2}}
\nn
&=&
2\pi \lk 1+\alpha \rk 
\alpha^{\frac{1}{2}}
\beta^{\frac{3}{2}}
\gamma. 
\eeq

\section{The ground-state energy in the second-order perturbation theory} 
In this appendix, 
we briefly show the derivation of the ground-state energy (\ref{E2nd})
obtained in the second-order perturbation theory.
The Fr{\"o}hlich-type Hamiltonian (\ref{FH1}) 
in the full second-quantized form  
is represented as $\mathcal{H}= \mathcal{H}_0+\mathcal{V}$, 
where 
the non-perturbative and perturbative parts, 
$\mathcal{H}_0$ and $\mathcal{V}$ 
are defined as,
\begin{eqnarray}
\mathcal{H}_0 
&=& 
\sum_u E_u^I a_u^\dagger a_u +E^b_0 N_0 +\sum_{s\neq0} E_s^b b_s^\dagger b_s,  
\\ 
\mathcal{V}
&=&
gN_0 \sum_{u,u'} C_{00;uu'}a_u^\dagger a_{u'}
+g\sqrt{N_0} \sum_{s\neq0,u,u'}
\lk C_{0s;uu'} b_s + C_{s0;uu'}b_s^\dagger\rk a_u^\dagger a_{u'},  
\end{eqnarray}
where $a_u$ ($a_u^\dagger$) is the annihilation (creation) operator of impurity 
with the labels of the abbreviated form $u=(n,l,m)$, 
and, also, the ground state is represented by $u=0$.
The overlap integrals $C_{ss';uu'}$ of the wave functions are defined by
\beq
C_{ss';uu'} = \int_{\br} {\phi^b_s}^*(\br) \phi^b_{s'}(\br) {\phi^I_u}^*(\br) \phi^I_{u'}(\br). 
\eeq 
In the diagrammatic method of the perturbation theory, 
the ground state energy is obtained from the summation of the connected diagrams (Goldstone's theorem). 
Up to the second order of $g$ for $J=J_z=0$, 
it becomes
\begin{eqnarray}
\braket{\mathcal{H}}
&=&\braket{\Phi_0|\mathcal{H}_0|\Phi_0}
+\braket{\Phi_0|\mathcal{V}|\Phi_0}
+\sum_{i}\frac{\braket{\Phi_0|\mathcal{V}|i}\braket{i|\mathcal{V}|\Phi_0}}{\braket{\Phi_0|\mathcal{H}_0|\Phi_0}-\braket{i|\mathcal{H}_0|i}}
\nn 
&=&
E_0^f+E_0^bN_0+gN_0C_{00;00}
\nn
&&+\sum_{s\neq0, u}
\frac{ \left|\braket{0|b_s a_u \mathcal{V} a_0^\dagger |0}\right|^2}
{E_0^I+E_0^bN_0
-\braket{0|b_s a_u 
\lk \sum_{u'} E_{u'}^I a_{u'}^\dagger a_{u'} + E^b_0 N_0 +\sum'_{s'} E_{s'}^b b_{s'}^\dagger b_{s'}\rk a_u^\dagger b_s^\dagger|0}}
\nn
&&+\sum_{u\neq0}
\frac{ \left|\braket{0|a_u \mathcal{V} a_0^\dagger |0}\right|^2}
{E_0^I+E_0^bN_0
-\braket{0|a_u 
\lk \sum_{u'} E_{u'}^I a_{u'}^\dagger a_{u'} + E^b_0 N_0 +\sum'_{s'} E_{s'}^b b_{s'}^\dagger b_{s'}\rk a_u^\dagger |0}}
\nn
&=&
E_0^I+E_0^bN_0+gN_0C_{00;00}
-g^2 N_0\sum_{s\neq0, u}\frac{C_{0s;0u}C_{s0;u0}}{E_u^I-E_0^I+E_s^b}
-g^2 N_0^2 \sum_{u\neq0}
\frac{C_{00;0u}C_{00;u0}}{E_u^I-E_0^I}, 
\nn 
\label{eq:Ept}
\eeq
where the non-perturbative ground state are defined by
\begin{eqnarray}
\ket{\Phi_0}&=&a_0^\dagger \ket{0} 
\end{eqnarray}
with the Fock vacuum of excited bosons $\ket{0}$ 
(the condensed state of the lowest-energy boson), 
and the intermediate states $\ket{i}$ are 
\begin{eqnarray}
\ket{i}&=& \ltk a_u^\dagger b_{s\neq0}^\dagger \ket{0}, 
\, a_{u\neq0}^\dagger \ket{0}\rtk. 
\end{eqnarray}
In order to make a fair comparison with the variational method, 
in the ground-state energy formula ($J=0$) , 
we take the impurity intermediate states 
up to $u=(1,0,0)$, 
and those of bosons only for $l=m=0$ 
(consistent with the $J=0$ state). 
Then we obtain the ground-state energy in the second-order perturbation theory:
\beq
\braket{\mathcal{H}} 
&\simeq&
E_{00}^I+E_{00}^bN_0+g\frac{N_0}{4\pi} H_{000}^0 
\nn
&-& g^2 \frac{N_0}{\lk 4\pi\rk^2}
\sum_{n\neq 0}
\lk 
\frac{|H_{n00}^c|^2}{E_{10}^I-E_{00}^I+E_{n0}^b}
+ \frac{|H_{n00}^0|^2}{E_{n0}^b}
\rk 
-g^2 \lk\frac{N_0}{4\pi}\rk^2 
\frac{|H_{000}^c|^2}{E_{10}^I-E_{00}^I},
\nn 
\eeq
which is just the eq.~ (\ref{E2nd}).
\end{document}